\input harvmac
\input xymatrix\input xyarrow
%\draftmode
\noblackbox

\def\abstract#1{
\vskip .5in\vfil\centerline
{\bf Abstract}\penalty1000
{{\smallskip\ifx\answ\bigans\leftskip 1pc \rightskip 1pc 
\else\leftskip 1pc \rightskip 1pc\fi
\noindent \abstractfont  \baselineskip=12pt
{#1} \smallskip}}
\penalty-1000}
\baselineskip=15pt plus 2pt minus 0pt
%%%%%%%%%%%%

\def\hth/#1#2#3#4#5#6#7{{\tt hep-th/#1#2#3#4#5#6#7}}
\def\nup#1({Nucl.\ Phys.\ $\us {B#1}$\ (}
\def\plt#1({Phys.\ Lett.\ $\us  {B#1}$\ (}
\def\cmp#1({Comm.\ Math.\ Phys.\ $\us  {#1}$\ (}
\def\prp#1({Phys.\ Rep.\ $\us  {#1}$\ (}
\def\prl#1({Phys.\ Rev.\ Lett.\ $\us  {#1}$\ (}
\def\prv#1({Phys.\ Rev.\ $\us  {#1}$\ (}
\def\mpl#1({Mod.\ Phys.\ Let.\ $\us  {A#1}$\ (}
\def\atmp#1({Adv.\ Theor.\ Math.\ Phys.\ $\us  {#1}$\ (}
\def\ijmp#1({Int.\ J.\ Mod.\ Phys.\ $\us{A#1}$\ (}
\def\jhep#1({JHEP\ $\us {#1}$\ (}

\def\bb#1{{\bar{#1}}}
\def\bx#1{{\bf #1}}
\def\cx#1{{\cal #1}}
\def\tx#1{{\tilde{#1}}}
\def\hx#1{{\hat{#1}}}
\def\vx#1{\vec{#1}}
\def\rmx#1{{\rm #1}}
\def\us#1{\underline{#1}}
\def\fc#1#2{{#1\over #2}}
\def\frac#1#2{{#1\over #2}}
\def\Psiv#1{{\vx \Psi}^{(#1)}}
\def\Phiv#1{{\vx \Phi}^{(#1)}}

\def\br{\hfill\break}
\def\noi{\noindent}

\def\al{\alpha}\def\be{\beta}\def\om{\omega}
\def\p{\partial}
\def\IP{\bf P}
\def\CY{Calabi--Yau }
%%%%%%%%%%%%%%%%ddd

\def\Gah{{\hx\Gamma}}
\def\zh{{\hx z}}

\def\lra#1{\matrix{#1\cr\longrightarrow\cr\phantom{1}}}
\def\bra#1{|#1\rangle_{RR}}

\def\Om{\Omega}\def\vphi{\varphi}\def\Si{\Sigma}
\def\p{\partial}
\def\IP{{\bf P}}\def\CC{{\bf C}}\def\ZZ{{\bf Z}}
\def\CY{Calabi--Yau }

\def\ph{\hat{\phi}}\def\th{\hat{t}}
\def\bb#1{\bar{#1}}

\def\rbra#1{|#1\rangle}
\def\ib{\bb i}
\def\flc{\nabla}\def\lefin{\end}
\def\zo{\hat{z}}\def\Ga{\Gamma}\def\Si{\Sigma}
\def\MM{\cx M _{_{\!\cx N=1}}}\def\Mm{\cx M _{_{\!\cx N=2}}}

\def\Roc{\cx R_{oc}}\def\Hom{\rmx{Hom}}
\def\delh{\hx \delta}
\def\doubref#1#2{\refs{{#1},{#2}}}
\def\ss{\scriptstyle}
\def\ti{\tilde}

\def\dd{\cdot}

\newif\ifnref

\def\doubref#1#2{\refs{{#1},{#2}}}

\nreffalse

%%%%%%%%%%%%%%%%%%%%%%%%%%%%%%%%%%%

\def\zo{{z_1}}

\def\xo{{t_1}}
\def\xs{{t_0}}

\def\IP{{\bf P}}
\def\CC{{\bf C}}
\def\cW{{\cal W}}
\def\cF{{\cal F}}
\def\xt(#1){\theta_{#1}}
\def\xa(#1){n_{#1}}
\def\xz(#1){z_{#1}}
\def\WZ{W_\Z}
\def\ll#1{l^{(#1)}}
\def\vx#1{\vec{#1}}
\def\GMC{{\cal A}}
\ifx\answ\bigans\else\fi
\def\xd{{\cdot}}

\def\Wt{{\tW}}
\def\tW{\widetilde{W}}
\def\two{2}
\def\zt{z_{\two}}
\def\xs{{t_\two}}
\def\WZ{\Wt_\Z}
\def\Thetap{\Theta^{+1}}

%cycle conventions:
\def\Z{Y}\def\z{z}
\def\Y{B}
\def\C{{B}}
\def\weqn#1{\xdef #1{(\noexpand\hyperref{}%
{equation}{\secsym\the\meqno}%
{\secsym\the\meqno})}\eqno(
%\hyperdef\hypernoname{equation}%
{\secsym\the\meqno}{\secsym\the\meqno})\eqlabeL#1%
\writedef{#1\leftbracket#1}\global\advance\meqno by1}
\def\weqnalign#1{\xdef #1{\noexpand\hyperref{}{equation}%
{\secsym\the\meqno}{(\secsym\the\meqno)}}%
\writedef{#1\leftbracket#1}%
%\hyperdef\hypernoname{equation}%
{\secsym\the\meqno}{\e@tf@ur#1}\eqlabeL{#1}%
\global\advance\meqno by1}
\def\abstract#1{
\vskip .5in\vfil\centerline
{\bf Abstract}\penalty1000
{{\smallskip\ifx\answ\bigans\leftskip 1pc \rightskip 1pc
\else\leftskip 1pc \rightskip 1pc\fi
\noindent \abstractfont  \baselineskip=12pt
{#1} \smallskip}}
\penalty-1000}
%%%%%%%%%%%%%%%%%%%%%%%%%%%%%

\def\eprt#1{{#1}}
\def\nihil#1{{\sl #1}}
\def\br{\hfill\break}

\lref\Grif{P.\ Griffiths, \nihil{
On the periods of certain rational integrals. I, II,} 
Ann. of Math. {\bf 90} (1969) 460,496.}

\lref\gkz{I.\ Gel'fand, M.\ Kapranov and A.\ Zelevinsky,
\nihil{Hypergeometric functions and toric varieties}, 
Funct. Anal. Appl. {\bf 23} no. 2 (1989), 12;\br
V.~V.~Batyrev and D.~van Straten,
\nihil{Generalized Hypergeometric Functions And Rational Curves
On Calabi-Yau Complete Intersections},
Commun.\ Math.\ Phys.\ {\bf 168}, 493 (1995).}

\lref\MiBo{See e.g.:
\nihil{Essays on mirror manifolds}, (S.\ Yau, ed.),
International Press 1992;
\nihil{Mirror symmetry II}, (B.\ Greene et al, eds.),
International Press 1997.}

\lref\Let{W.\ Lerche, P.\ Mayr and N.\ Warner, 
\nihil{Holomorphic $N=1$ Special Geometry 
of Open--Closed Type II Strings,}
\eprt{hep-th/0207259}. 
%%CITATION = HEP-TH 0207259;%%
}

\lref\Guk{S.~Gukov,
\nihil{Solitons, superpotentials and calibrations,}
Nucl.\ Phys.\ B {\bf 574}, 169 (2000),
hep-th/9911011.}

\lref\GVW{
S.\ Gukov, C.\ Vafa and E.\ Witten, 
\nihil{CFT's from Calabi-Yau four-folds,}
 Nucl.\ Phys.\ B{\bf 584} 69 (2000), 
Erratum-ibid.\ B{\bf 608} 477 (2001),
\eprt{hep-th/9906070}.
%%CITATION = HEP-TH 9906070;%%
}

\lref\AVi{M.\ Aganagic and C.\ Vafa, 
\nihil{Mirror symmetry, D-branes and counting holomorphic discs,}
\eprt{hep-th/0012041}. 
%%CITATION = HEP-TH 0012041;%%
}

\lref\AVii{M.\ Aganagic, A.\ Klemm and C.\ Vafa, 
\nihil{Disk instantons, mirror symmetry and the duality web,}
 Z.\ Naturforsch.\ A{\bf 57} 1 (2002), 
\eprt{hep-th/0105045}. 
%%CITATION = HEP-TH 0105045;%%
}

\lref\BCOV{M.\ Bershadsky, S.\ Cecotti, H.\ 
Ooguri and C.\ Vafa, 
\nihil{Kodaira-Spencer theory of gravity and exact 
results for quantum string amplitudes,}
 Commun.\ Math.\ Phys.\ {\bf 165} 311 (1994), 
\eprt{hep-th/9309140}. 
%%CITATION = HEP-TH 9309140;%%
}

\lref\Cac{
{F.\ Cachazo, K.\ A.\ Intriligator and C.\ Vafa, 
\nihil{A large N duality via a geometric transition,}
 Nucl.\ Phys.\ B{\bf 603} 3 (2001),
\eprt{hep-th/0103067}; 
%%CITATION = HEP-TH 0103067;%%
}\br
J.~D.~Edelstein, K.~Oh and R.~Tatar,
\nihil{Orientifold, geometric transition and large 
N duality for SO/Sp gauge  theories,}
JHEP {\bf 0105}, 009 (2001), hep-th/0104037;\br
%%CITATION = HEP-TH 0104037;%%
{F.\ Cachazo, S.\ Katz and C.\ Vafa, 
\nihil{Geometric transitions and N = 1 quiver theories,}
\eprt{hep-th/0108120}; 
%%CITATION = HEP-TH 0108120;%%
}\br
{F.\ Cachazo, B.\ Fiol, K.\ A.\ Intriligator, 
S.\ Katz and C.\ Vafa, 
\nihil{A geometric unification of dualities,}
 Nucl.\ Phys.\ B{\bf 628} 3 (2002),
\eprt{hep-th/0110028}. 
%%CITATION = HEP-TH 0110028;%%
}
}

\lref\Can{P.\ Candelas, X.\ C.\ De La Ossa, P.\ S.\ Green 
and L.\ Parkes, 
\nihil{A Pair Of Calabi-Yau Manifolds As An Exactly 
Soluble Superconformal Theory,}
 Nucl.\ Phys.\ B{\bf 359} 21 (1991). 
%%CITATION = NUPHA,B359,21;%%
}

\lref\SGnPF{A.\ Ceresole, R.\ D'Auria, S.\ Ferrara, 
W.\ Lerche and J.\ Louis, 
\nihil{Picard-Fuchs equations and special geometry,}
 Int.\ J.\ Mod.\ Phys.\ A{\bf 8} 79 (1993),
\eprt{hep-th/9204035}. 
%%CITATION = HEP-TH 9204035;%%
}

\lref\Detal{I.\ Brunner, M.\ R.\ Douglas, A.\ E.\ Lawrence 
and C.\ R\"omelsberger, 
\nihil{D-branes on the quintic,}
 JHEP{\bf 0008} 015 (2000),
\eprt{hep-th/9906200}. 
%%CITATION = HEP-TH 9906200;%%
}

\lref\GMP{B.\ R.\ Greene, D.\ R.\ Morrison and M.\ R.\ Plesser, 
\nihil{Mirror manifolds in higher dimension,}
 Commun.\ Math.\ Phys.\ {\bf 173} 559 (1995),
\eprt{hep-th/9402119}. 
%%CITATION = HEP-TH 9402119;%%
}

\lref\HMA{
{C.\ Hofman and W.\ K.\ Ma, 
\nihil{Deformations of topological open strings,}
 JHEP{\bf 0101} 035 (2001),
\eprt{hep-th/0006120}. 
%%CITATION = HEP-TH 0006120;%%
}}

\lref\HIV{K.\ Hori, A.\ Iqbal and C.\ Vafa, 
\nihil{D-branes and mirror symmetry,}
\eprt{hep-th/0005247}. 
%%CITATION = HEP-TH 0005247;%%
}

\lref\HV{K.\ Hori and C.\ Vafa, 
\nihil{Mirror symmetry,}
\eprt{hep-th/0002222}. 
%%CITATION = HEP-TH 0002222;%%
}

\lref\KKLM{{S.\ Kachru, S.\ Katz, A.\ E.\ Lawrence and J.\ McGreevy, 
\nihil{Open string instantons and superpotentials,}
 Phys.\ Rev.\ D{\bf 62} 026001 (2000),
\eprt{hep-th/9912151}; 
%%CITATION = HEP-TH 9912151;%%
}{
 \nihil{Mirror symmetry for open strings,}
 Phys.\ Rev.\ D{\bf 62} 126005 (2000),
\eprt{hep-th/0006047}. 
%%CITATION = HEP-TH 0006047;%%
}}

\lref\KLMVW{A.\ Klemm, W.\ Lerche, P.\ Mayr, C.\ Vafa and N.\ P.\ Warner, 
\nihil{Self-Dual Strings and N=2 Supersymmetric Field Theory,}
 Nucl.\ Phys.\ B{\bf 477} 746 (1996), 
\eprt{hep-th/9604034}. 
%%CITATION = HEP-TH 9604034;%%
}

\lref\LVW{
W.~Lerche, C.~Vafa and N.~P.~Warner,
\nihil{Chiral Rings In N=2 Superconformal Theories,}
Nucl.\ Phys.\ B {\bf 324}, 427 (1989).
%%CITATION = NUPHA,B324,427;%%
}

\lref\LSW{W.~Lerche, D.~J.~Smit and N.~P.~Warner,
\nihil{
Differential equations for periods and flat coordinates in two-dimensional topological matter theories,}
Nucl.\ Phys.\ B {\bf 372}, 87 (1992) hep-th/9108013.}

\lref\LM{W.\ Lerche and P.\ Mayr, 
\nihil{On N = 1 mirror symmetry for open type II strings,}
\eprt{hep-th/0111113}. 
%%CITATION = HEP-TH 0111113;%%
}

\lref\PMsusy{P.\ Mayr, 
\nihil{On supersymmetry breaking in string theory and 
its realization in brane worlds,}
 Nucl.\ Phys.\ B{\bf 593} 99 (2001), 
\eprt{hep-th/0003198}. 
%%CITATION = HEP-TH 0003198;%%
}

\lref\PM{P.\ Mayr, 
\nihil{$N=1$ mirror symmetry and open/closed string duality,}
Adv.\ Theor.\ Math.\ Phys.\  {\bf 5}, 213 (2001),
\eprt{hep-th/0108229}.}

\lref\PMffs{P.\ Mayr,
\nihil{Mirror symmetry, $N = 1$ superpotentials and
 tensionless strings on Calabi-Yau four-folds,}
 Nucl.\ Phys.\ B{\bf 494} 489 (1997), 
\eprt{hep-th/9610162}. 
%%CITATION = HEP-TH 9610162;%%
}

\lref\OV{H.~Ooguri and C.~Vafa,
\nihil{Knot invariants and topological strings,}
Nucl.\ Phys.\ B {\bf 577}, 419 (2000), hep-th/9912123.}

\lref\OOY{H.\ Ooguri, Y.\ Oz and Z.\ Yin, 
\nihil{D-branes on Calabi-Yau spaces and their mirrors,}
 Nucl.\ Phys.\ B{\bf 477} 407 (1996),
\eprt{hep-th/9606112}. 
%%CITATION = HEP-TH 9606112;%%
}

\lref\TV{T.\ R.\ Taylor and C.\ Vafa, 
\nihil{RR flux on Calabi-Yau and partial supersymmetry breaking,}
 Phys.\ Lett.\ B{\bf 474} 130 (2000), 
\eprt{hep-th/9912152}. 
%%CITATION = HEP-TH 9912152;%%
}

\lref\CVlargeN{C.\ Vafa, 
\nihil{Superstrings and topological strings at large N,}
 J.\ Math.\ Phys.\ {\bf 42} 2798 (2001),
\eprt{hep-th/0008142}. 
%%CITATION = HEP-TH 0008142;%%
}

\lref\vafaext{C.\ Vafa, 
\nihil{Extending mirror conjecture to Calabi-Yau with bundles,}
\eprt{hep-th/9804131}. 
%%CITATION = HEP-TH 9804131;%%
}

\lref\Witlsm{E.\ Witten, 
\nihil{Phases of N = 2 theories in two dimensions,}
 Nucl.\ Phys.\ B{\bf 403} 159 (1993),
\eprt{hep-th/9301042};\br
D.~R.~Morrison and M.~Ronen Plesser,
\nihil{Summing the instantons: Quantum cohomology and 
mirror symmetry in toric varieties,}
Nucl.\ Phys.\ B {\bf 440}, 279 (1995)
hep-th/9412236.}

\lref\WQCD{E.\ Witten, 
\nihil{Branes and the dynamics of QCD, }
 Nucl.\ Phys.\ B{\bf 507} 658 (1997),
\eprt{hep-th/9706109}. 
%%CITATION = HEP-TH 9706109;%%
}

\lref\WitCS{E.\ Witten, 
\nihil{Chern-Simons gauge theory as a string theory,}
\eprt{hep-th/9207094}. 
%%CITATION = HEP-TH 9207094;%%
}
 
\lref\wittop{E.\ Witten, 
\nihil{Mirror manifolds and topological field theory,}
\eprt{hep-th/9112056}; 
%%CITATION = HEP-TH 9112056;%%
}  

\lref\AVo{
{M.\ Aganagic and C.\ Vafa, 
\nihil{Mirror symmetry and a G(2) flop,}
\eprt{hep-th/0105225}; \br
%%CITATION = HEP-TH 0105225;%%
}{ 
\nihil{G(2) manifolds, mirror symmetry and geometric engineering,}
\eprt{hep-th/0110171}.
%%CITATION = HEP-TH 0110171;%%
}}
\lref\govi{S.\ Govindarajan, T.\ Jayaraman and T.\ Sarkar, 
\nihil{Disc instantons in linear sigma models,}
\eprt{hep-th/0108234}.
%%CITATION = HEP-TH 0108234;%%
}
\lref\IK{A.\ Iqbal and A.\ K.\ Kashani-Poor, 
\nihil{Discrete symmetries of the superpotential and 
caculation of disk invariants,}
\eprt{hep-th/0109214}.
%%CITATION = HEP-TH 0109214;%%
}
\lref\AAHV{B.\ Acharya, M.\ Aganagic, K.\ Hori and C.\ Vafa, 
\nihil{Orientifolds, mirror symmetry and superpotentials,}
\eprt{hep-th/0202208}.
%%CITATION = HEP-TH 0202208;%%
}
\lref\DFG{D.\ E.\ Diaconescu, B.\ Florea and A.\ Grassi, 
\nihil{Geometric transitions and open string instantons,}
\eprt{hep-th/0205234}. 
%%CITATION = HEP-TH 0205234;%%
}

\lref\LiSG{
J.~Li and Y.~S.~Song,
\nihil{Open string instantons and relative stable morphisms,}
Adv.\ Theor.\ Math.\ Phys.\  {\bf 5}, 67 (2002)
hep-th/0103100.
}
\lref\KatzVM{
S.~Katz and C.~C.~Liu,
\nihil{Enumerative Geometry of Stable Maps with Lagrangian Boundary Conditions and Multiple Covers of the Disc,}
Adv.\ Theor.\ Math.\ Phys.\  {\bf 5}, 1 (2002)
math.ag/0103074.
}
\lref\GZ{T.\ Graber and E.\ Zaslow, 
\nihil{Open string Gromov-Witten invariants: 
Calculations and a mirror 'theorem',}
 \eprt{hep-th/0109075}.}

\lref\yaip{P.\ Mayr, 
\nihil{Summing up open string instantons 
and $N=1$ string amplitudes,}
\eprt{hep-th/0203237}. 
}

\lref\CS{
J.~M.~Labastida and M.~Marino,
\nihil{Polynomial invariants for torus knots and topological strings,}
Commun.\ Math.\ Phys.\  {\bf 217}, 423 (2001), hep-th/0004196;\br
P.~Ramadevi and T.~Sarkar,
\nihil{On link invariants and topological string amplitudes,}
Nucl.\ Phys.\ B {\bf 600}, 487 (2001), hep-th/0009188;\br
J.~M.~Labastida, M.~Marino and C.~Vafa,
\nihil{Knots, links and branes at large N,}
JHEP {\bf 0011}, 007 (2000), hep-th/0010102;\br
J.~M.~Labastida and M.~Marino,
\nihil{A new point of view in the theory of knot and link invariants,}
math.qa/0104180;\br
M.~Marino and C.~Vafa,
\nihil{Framed knots at large N,}
hep-th/0108064;\br
D.\ E.\ Diaconescu, B.\ Florea and A.\ Grassi, 
\nihil{Geometric Transitions, del Pezzo Surfaces and
open String Instantons,}
\eprt{hep-th/0206163};\br
{M.\ Aganagic, M.\ Marino and C.\ Vafa, 
\nihil{All Loop Topological String Amplitudes From Chern-Simons Theory,}
\eprt{hep-th/0206164}. 
%%CITATION = HEP-TH 0206164;%%
}}

\lref\sgcy{
E.~Cremmer, C.~Kounnas, A.~Van Proeyen, J.~P.~Derendinger, 
S.~Ferrara, B.~de Wit and L.~Girardello,
\nihil{Vector Multiplets Coupled To N=2 Supergravity: 
Superhiggs Effect, Flat Potentials And Geometric Structure,}
Nucl.\ Phys.\ B {\bf 250}, 385 (1985);\br
%%CITATION = NUPHA,B250,385;%%
B.~de Wit, P.~G.~Lauwers and A.~Van Proeyen,
\nihil{Lagrangians Of N=2 Supergravity - Matter Systems,}
Nucl.\ Phys.\ B {\bf 255}, 569 (1985);\br
%%CITATION = NUPHA,B255,569;%%
A.~Strominger,
\nihil{Special Geometry,}
Commun.\ Math.\ Phys.\  {\bf 133}, 163 (1990).}

\lref\Cec{S.~Cecotti,
\nihil{Geometry Of N=2 Landau-Ginzburg Families,}
Nucl.\ Phys.\ B {\bf 355}, 755 (1991);
%%CITATION = NUPHA,B355,755;%%
\nihil{N=2 Landau-Ginzburg Versus Calabi-Yau Sigma Models: 
Nonperturbative Aspects,}
Int.\ J.\ Mod.\ Phys.\ A {\bf 6}, 1749 (1991).
%%CITATION = IMPAE,A6,1749;%%
}

\lref\CeVa{S.\ Cecotti and C.\ Vafa, 
\nihil{Topological--antitopological fusion,}
 Nucl.\ Phys.\ B{\bf 367} 359 (1991). 
%%CITATION = NUPHA,B367,359;%%
}

\lref\newsugra{J.~Polchinski and A.~Strominger,
\nihil{New Vacua for Type II String Theory,}
Phys.\ Lett.\ B {\bf 388}, 736 (1996)
hep-th/9510227;\br
J.~Michelson,
\nihil{Compactifications of type IIB strings to four dimensions with  non-trivial classical potential,}
Nucl.\ Phys.\ B {\bf 495}, 127 (1997)
hep-th/9610151;\br
G.~Curio, A.~Klemm, D.~L\"ust and S.~Theisen,
\nihil{On the vacuum structure of type II string compactifications on  Calabi-Yau spaces with H-fluxes,}
Nucl.\ Phys.\ B {\bf 609}, 3 (2001)
hep-th/0012213;\br
L.~Andrianopoli, R.~D'Auria and S.~Ferrara,
\nihil{Consistent reduction of N = 2 $\to$ N = 1 four dimensional supergravity  coupled to matter,}
Nucl.\ Phys.\ B {\bf 628}, 387 (2002)
hep-th/0112192;\br
L.~Andrianopoli, R.~D'Auria, S.~Ferrara and M.~A.~Lledo,
\nihil{Super Higgs effect in extended supergravity,}
hep-th/0202116;\br
J.~Louis and A.~Micu,
\nihil{Type II theories compactified on Calabi-Yau threefolds in the presence  of background fluxes,}
Nucl.\ Phys.\ B {\bf 635}, 395 (2002)
hep-th/0202168;\br
J.~Louis,
\nihil{Aspects of spontaneous N = 2 $\to$ N = 1 breaking in supergravity,}
hep-th/0203138.
}

\lref\MHV{P. Deligne, \nihil{Theori\'e de Hodge I-III},
Actes de Congr\`es international de Mathematiciens (Nice,1970),
Gauthier-Villars {\bf 1} 425 (1971);
Publ. Math. IHES {\bf 40} 5 (1971);
Publ. Math. IHES {\bf 44} 5 (1974);\br
J. Carlson, M. Green, P. Griffiths and J. Harris,
\nihil{Infinitesimal variations of 
Hodge structure (I)}, Compositio Math. {\bf 50} (1983), no. 2-3, 109.}

\lref\Grifdeq{P.A. Griffiths, 
\nihil{A theorem concerning the differential equations satisfied
by normal functions associated to algebraic cycles},
Am. J. Math {\bf 101} 96 (1979).}

\lref\Kar{M. Karoubi and C. Leruste, 
\nihil{Algebraic topology
via differential geometry}, Cambridge Univ. Press,
Cambridge, 1987.}

\lref\Griii{P.A. Griffiths, \nihil{Some results on algebraic cycles
in algebraic manifolds}, Algebraic Geometry (Internat. Colloq., Tata Inst. 
Fund. Res., Bombay, 1968), 93--191, 1969, Oxford Univ. Press, London.}

%%%%%%%%%%%%%%%%%%%
\vskip-2cm
\Title{\vbox{
\rightline{\vbox{\baselineskip12pt\hbox{CERN-TH/2002-175}
         \hbox{hep-th/0208039}}}\vskip0cm}}
{$N=1$ Special Geometry,}\vskip -28pt
\centerline{\titlefont Mixed Hodge Variations and Toric Geometry}
\vskip 20pt

\abstractfont 
\centerline{W. Lerche$^\flat$, P. Mayr$^\flat$ and N. Warner$^\sharp$}

\vskip 0.4cm
\centerline{$^\flat$ \ninepoint
CERN Theory Division, CH-1211 Geneva 23, Switzerland}
\centerline{$^\sharp$ \ninepoint Department of Physics and Astronomy,
University of Southern California, }
\centerline{\ninepoint Los Angeles, CA 90089-0484, USA}
\vskip -0.0cm
\abstract{
We study the superpotential of a certain class of $\cx N=1$
supersymmetric type II compactifications with fluxes and $D$-branes.
We show that it has an important two-dimensional
meaning in terms of a chiral ring of the topologically twisted
theory on the world-sheet. In the open-closed string B-model, this
chiral ring is isomorphic to a certain relative cohomology group
$V$,  which is the appropriate mathematical concept to deal with 
both the open and closed string sectors. 
The family of mixed Hodge structures on $V$ then implies for the
superpotential to have a certain geometric structure. This structure
represents a holomorphic, $\cx N=1$ supersymmetric generalization
of the well-known $\cx N=2$ special geometry.
It defines an integrable connection on the topological family of open-closed B-models, and a set of special coordinates on the space $\cal M$ of vev's in $\cx N=1$ chiral multiplets.
We show that it can be given a very concrete and simple realization
for linear sigma models, which leads to a powerful and systematic
method for computing the exact non-perturbative $\cx N=1$
superpotentials for a broad class of toric $D$-brane geometries.
}
\vskip1cm
\Date{\vbox{\hbox{ {August 2002}}
}}
\goodbreak

%%%%%%%%%%%%%%%%%%%%%%%%%%%%%%%%%%%%%%%%%%%%%%%%%
\newsec{Introduction}
%%%%%%%%%%%%%%%%%%%%%%%%%%%%%%%%%%%%%%%%%%%%%%%%%

Mirror symmetry \MiBo\ has proven to be a most valuable tool
for doing exact computations in $\cx  N=2$ supersymmetric string theories.
The geometric methods provide insights into the non-perturbative dynamics of
such theories and, at a more conceptual level, they provide a 
glimpse of what one may call ``stringy quantum geometry.'' The underlying
concept is that of the 2d topological field theory (TFT) on the 
string world-sheet, which relates non-perturbative space-time effects
to computable geometric data of the compactification manifold.

While much of the work done in the past has been
concerned with \CY compactifications of the type II string
with $\cx N=2$ supersymmetry, methods have recently been developed
for dealing with $\cx N=1$ supersymmetric\foot{What is meant here is 
that the 4d theory can be described by an effective $\cx N=1$ supergravity
action. Supersymmetry may, and in many cases will, 
be broken by the presence of a superpotential.} compactifications as well. 
They apply in particular to open-closed type II string compactifications 
on \CY manifolds $X$, having extra fluxes on cycles in $X$, and/or extra 
branes wrapped on cycles in $X$ and filling space-time.
Most notably,
generalizations of mirror symmetry  have been used to 
compute exact 4d $\cx N =1$ superpotentials,
including all corrections from world-sheet instantons of 
sphere \TV-\phantom{\PMsusy\CVlargeN}\hskip-30pt \ \ \Cac\
and disc 
{\WitCS-\phantom{\vafaext\KKLM\AVi\AVii\PM\govi\AVo\IK\LM\AAHV\DFG}
\hskip-163pt\ \Let}
topologies.
Instanton corrections to the superpotential and to other 
4d $\cx N=1$ amplitudes such as the gauge kinetic
$F$-terms, have been also computed by
localization techniques \LiSG-\phantom{\KatzVM\GZ}\hskip-30pt\yaip\ 
and Chern-Simons theory \doubref\OV\CS.

The purpose of this note, as well as of the earlier companion paper \Let, 
is to study the geometry of
the superpotential on the space $\MM$ of scalar vev's in $\cx N=1$ 
chiral multiplets. These F-terms carry an interesting geometric structure 
inherited from the underlying 
2d topological field theory on the world-sheet. 
This ``holomorphic $\cx N=1$ special geometry'' is
a close relative of the well-known $\cx N=2$ 
special geometry \doubref\sgcy\BCOV\
of the moduli space of type II compactifications 
without fluxes and branes. However, the geometry of the 
$\cx N=1$ F-terms inherited from the TFT is {\it not} a consequence
of space-time supersymmetry; rather it should be interpreted as a
distinguished feature of the string effective supergravity, as opposed 
to a generic supergravity theory.

The study of the holomorphic\foot{There
is also a non-holomorphic part given by the D-terms,
most notably the K\"ahler potential for the chiral multiplets. 
While this would be 
interesting to study as well, perhaps in the framework of the $tt^*$
equations of \doubref\CeVa\BCOV, this is beyond the scope of the paper.}
$\cx N=1$ special geometry
has been started in \doubref\PM\LM\ in the context of 
a certain open-closed string duality, which
relates the $\cx N=1$ special geometry to the geometry of 
the moduli space of \CY 4-folds. In \Let\ we have outlined
how the $\cx N=1$ special geometry arises 
as the consequence of systematically incorporating fluxes
and branes into the familiar ideas and methods of 
mirror symmetry for \CY 3-folds. In the present paper
we complete the discussion of \Let\ in two respects.

First, we will develop a geometric representation of 
the chiral ring and the integrable connection $\nabla$
on the (B-twisted) family of TFT's over the $\cx N=1$
deformation space $\MM$. Specifically, we show that
the TFT concepts are equivalent to the mixed Hodge variation defined 
on a certain relative cohomology group 
associated with the flux and brane geometry. 
The fundamental r\^ole of the variation of 
the Hodge structure in the construction of $\cx N=2$ mirror symmetry
is therefore taken over by the {\it mixed} Hodge structure
on the relative cohomology group. 
In particular, a system of differential equations associated
with the Hodge structure gives a powerful and systematic 
technique for computing $i)$ a set of special, flat 
coordinates $t_A$ on $\MM$, and $ii)$ exact expressions for
the holomorphic
potentials $\cx W_K$ of $\cx N=1$ special geometry.
These potentials $\cx W_K$ represent the basic topological data of the theory,
and are in a quite precise sense the $\cx N=1$ analogs of
the familiar $\cx N=2$ prepotential $\cF$. Moreover,
the holomorphic potentials $\cx W_K$ have an important physical meaning,  
in that they describe the space-time instanton corrected 
superpotential of the effective $\cx N=1$ supergravity in 
four dimensions.

Second, we will develop an explicit description of
the relative cohomology and its mixed Hodge structure
in terms of linear sigma models, or, equivalently,
of toric geometry. Just like for the closed string, 
the toric formulation is highly effective also for the open-closed string compactification, and leads to a straightforward calculation
of the special coordinates $t_A$ (by the $\cx N=1$ mirror map) 
and of the
potentials $\cx W_K$. More specifically, the problem to compute the 
space-time instanton expansion to the four-dimensional superpotential is 
essentially reduced to finding an appropriate 
generalized hypergeometric series that solves a certain
system of linear differential equations associated with the toric
$D$-brane geometry.

The organization of this note is as follows. 
Sect.~2 contains an overview over the main ideas
and results that were outlined in \Let\ 
and will be further developed in the present paper.
In sect.~3, we discuss the 2d chiral ring of the open-closed
B-model, which is an extension of the bulk chiral ring by boundary operators. 
We identify the chiral ring elements with sections
of a certain relative cohomology group, $H^3(X,\Z)$.
In sect.~4 we consider continuous families of topological 
B-models over the space $\MM$ of the open-closed string deformations $z_A$,
which represent scalars vev's of 4d chiral $\cx N=1$ multiplets.
The central objects are the variation of the 
mixed Hodge structure on the relative cohomology group and  
an integrable connection $\nabla$
which is constructed as the Gauss-Manin connection on the
relative cohomology bundle. 

In sect.~5 we elaborate on the powerful 
methods of toric geometry to study the relative cohomology and 
mixed Hodge structure for open-closed compactifications described by linear sigma models. The system of differential equations associated with
$\nabla$ takes a simple, generalized hypergeometric form. 
Accordingly, its solutions,
which represent the $\cx N=1$ mirror map and the holomorphic potentials
$\cx W_K$, are generalizations of hypergeometric series.
Moreover, a certain ambiguity for non-compact geometries, the
so-called framing ambiguity discovered in \AVii,
is identified with the degree of a certain hypersurface 
$\Z$ that defines the relative cohomology $H^3(X,\Z)$. 
Finally, in sect.~6
we illustrate the toric methods at the hand of a detailed sample
computation.

%%%%%%%%%%%%%%%%%%%%%%%%%%%%%%%%%%%%%%%%%%%%%%%%%
\newsec{Overview}
%%%%%%%%%%%%%%%%%%%%%%%%%%%%%%%%%%%%%%%%%%%%%%%%%

Recall that  the vector multiplet moduli space $\Mm$ of the type II
string compactified on a  \CY manifold $X$ to four dimensions is a 
restricted K\"ahler manifold with so-called $\cx N=2$ special geometry
\doubref\sgcy\BCOV.  The $\cx N=2$ supersymmetric effective action
for the vector multiplets is defined by a holomorphic prepotential
$\cx F(z_a)$, where $z_a$ are generic coordinates on the space $\Mm$ of
scalar vev's.  The special geometry of $\Mm$ can be traced back to 
properties of the underlying topological field theory on the string
world-sheet, which leads to beautiful connections between physics and
geometry.  Indeed the key concepts of the
geometry can be directly identified with properties of the internal 
twisted TFT \BCOV.  A central object of this TFT is the ring of 
two-dimensional, primary chiral superfields \LVW.
Its moduli dependent structure constants, or operator product coefficients,
defined by\foot
{Here $t_a=t_a(z_b)$ denote the special, topological coordinates
that will be discussed in detail later. Moreover $\p_a=\p/\p t_a$.
} 
\eqn\ccr{
\cx R_{cl}:\ \ \phi_a\cdot \phi_b={C_{ab}}^{\ c}(t_\dd)\,
\phi_c\, ,}
are related to three-point functions which are
given by derivatives of the prepotential:\eqn\cscr{ {C_{ab}}^{\ c}(t_\dd)=\cx F_{abc}(t_\dd)~=~\p_a\p_b\p_c\cx
F(t_\dd).}
The prepotential, and thus the 
effective action for the vector multiplets, can be
recovered from the chiral ring coefficients $C_{ab}^{\ \ c}(t_\dd)$ through
integration.

In a certain variant of the 2d TFT, the so-called B-model \wittop,
there are no quantum corrections to the chiral ring, 
and the $C_{ab}^{\ \ c}(t_\dd)$ 
can be computed via classical geometry. They appear as particular
components of the period matrix for the elements of $H^3(X,\CC)$:
\eqn\pmi{
\Pi^\al_i = \langle \Ga^\al,\, \Phi_i\rangle = \int_{\Ga^\al}\, \Phi_i
=\pmatrix{
1&t_a&\cx F_a&2\cx F -t_a\cx F_a\cr
0&\delta_{ab}&\cx F_{ab}&\cx F_b-t_a \cx F_{ab}\cr
0&0          &\cx F_{abc}   &-t_a \cx F_{abc}\cr
0&0&...&...\cr}
.}

Here the forms $\Phi_i\in H^3(X,\CC)$ are, up to normalization, 
geometric representatives
for the elements of the chiral ring, and the $\{\Ga^\al\}$ 
constitute a fixed
basis for the homology 3-cycles, $H_3(X,\ZZ)$. 

The underlying geometric structure of the period integrals is the 
variation of Hodge structure on a family of \CY manifolds, whose
complex structures are parametrized by points in $\Mm$.
%If $X$ is defined as a hypersurface, 
%the complex structure is parametrized by the complex coefficients $a_j$ 
%in the defining polynomial, where the $a_j$ have a
%complicated functional relationship with the flat coordinates $t_a$. 
The integrable Gauss-Manin connection $\flc_a$ on the Hodge bundle with
fibers $H^3(X)$ leads to a 
system of differential equations in the moduli that enables one
to compute the period matrix $\Pi^\al_i$ from the solutions.
Thus, solving these differential equations provides a simple
way of computing the flat coordinates $t_a$ and the 
holomorphic prepotential $\cx F$ 
for a given threefold $X$.

Our purpose here is to derive similar results for 
$\cx N=1$  supersymmetric theories obtained from compactifying 
open-closed type II strings on a
\CY manifold $X$. In these theories,  
supersymmetry is reduced by the presence of 
fluxes of the closed string gauge potentials, and/or by the presence of an 
open string sector from background (D-)branes that
wrap cycles in $X$ and fill space-time. 
Specifically, we will mainly 
consider D5-branes of the type IIB string that
wrap certain 2-cycles in $X$. The latter are mapped by mirror
symmetry to type IIA D6-branes partially wrapped on 3-cycles.

A type IIB compactification with such branes and fluxes may be described
by an effective $\cx N=1$ supergravity with a non-trivial superpotential
on the space $\MM$ of vev's in $\cx N=1$ chiral multiplets.%
\foot{We will sometimes loosely refer to this space as the
``$\cx N=1$ moduli space'', despite of the fact that it is generically
obstructed by the superpotential $W(\phi)$ 
that lifts flat directions and may break supersymmetry 
spontaneously.  This notion is well motivated if 
the superpotential is completely non-perturbative, as is 
the case for the moduli of
branes wrapping supersymmetric cycles in $X$.}
This space comprises the complex structure 
moduli $z_a$ from the closed, and brane moduli $\zh_\al$ 
from the open string sector. 
As discussed in \Let,
the general superpotential on $\MM$, depending on both
the closed and open string moduli, 
can be written as 
\eqn\gensp{
W_{\cx N=1}\ =\ W_{cl}(z_a)+W_{op}(z_a,\zh_\al)\ =\ 
\sum_\Si N_\Si \, \Pi^\Si(z_A),
}
where 
\eqn\ZAdef{
z_A\ \equiv\ (z_a,\zh_\al),\qquad\ \ A\ =\ 
1,\dots,M\equiv{\rm dim}\, \MM,
} 
and where $\Pi^\Si$ is the {\it relative period vector}
\eqn\lpv{ \Pi^\Si(z_A) = \int_{\Ga^\Si} 
\Om\,, \qquad \Ga^\Si \in H_3(X,\Y;\ZZ). }
Here $\Ga^\Si\in H_3(X,\Y;\ZZ)$ is a fixed basis of relative 
homology 3-cycles  in $X$ with boundaries on the union of 2-cycles
$\Y=\cup_\nu \C_\nu$, where the $\C_\nu$ denote the 2-cycles 
wrapped by 5-branes. The elements of $H_3(X,\Y;\ZZ)$ are 
are $i)$ the familiar 3-cycles in $X$ without boundaries, whose volumes
specify the flux superpotential \refs{\TV,\PM,\GVW}, and $ii)$ the 3-chains
with boundaries on the 2-cycles $B_\nu$, whose 3-volumes 
govern the brane superpotential \refs{\WQCD,\AVi,\KKLM}.\foot{See 
also \newsugra\ for a further discussion of branes and fluxes in the effective 
supergravity.}
The vector $\Pi^\Si$ thus uniformly combines the period 
integrals of $\Om$ over 3-cycles $\Gamma^\al$, with 
integrals over the 3-chains $\Gah^\nu$ whose boundaries are wrapped by D-branes.%
\foot{More precisely,
the pairing between homology and cohomology 
is defined in relative cohomology, as discussed in 
sect.~4.2}
The coefficients $N_\Si$ in \gensp\ label the various 
flux and brane sectors. Specifically, $N_\Si=n_\Si + \tau\, m_\Si$,
where $n_\Si$ measures either the integral RR flux on $\Ga^\Si$ if $\Si$ labels
a 3-cycle, or the number of D5-branes if $\Si$ labels
a 3-chain. Similarly, the integers $m_\Si$ specify the NS fluxes and
branes. 

In fact, there is an alternative interpretation
of the 4d superpotential, in terms of the tension of a
BPS domain wall obtained by wrapping a 5-brane on a relative 3-cycle 
in $X$ \doubref\GVW\Guk\TV. The possible wrappings are determined by the
relative homology group $H_3(X,\Z;\ZZ)$. We are thus led to identify 
the integral relative cohomology group with the lattice of 4d BPS charges
(with two copies for the NS and RR sector, respectively) 
$$
\Gamma_{BPS} = H^3(X,\Z;\ZZ) \otimes (1\oplus \tau),
$$
where the charge $Q\in \Gamma_{BPS}$ is specified by the 
integers $n_\Si$ and $m_\Si$. The holomorphic
$\cx N=1$ special geometry of the string effective supergravity 
may thus also be interpreted as the BPS geometry 
of 4d domain walls with charges $Q\in\Ga_{BPS}$, very much as 
the $\cx N=2$ special geometry describes the 
BPS geometry of four-dimensional particles. 

Let us now briefly summarize our results on the underlying topological
structure of the superpotential, 
partly recapitulating statements made in \Let, and partly previewing
the discussion in the later sections of the present paper. 

A central object in the
2d TFT on the string world-sheet is the extended 
chiral ring of the open-closed type II theory:
\eqn\rocky{
\Roc:\qquad\phi_I \cdot \phi_J = C_{IJ}^{\ \ K}(z_A)\, \phi_K.}
This open-closed chiral ring is an extension of
the bulk chiral ring \ccr\ by boundary operators.
Its structure constants depend on 
both the open and closed string moduli $z_A$ parametrizing $\MM$. 
The 2d superfields $\phi_I$ form a basis for the local BRST 
cohomology of the 2d TFT.

As will be discussed, in the topological B-model 
the ring elements $\phi_I$ can be identified with elements of a 
certain relative cohomology group, $V=H^3(X,\Z)$; here 
$\Z$ is a union of hypersurfaces in $X$ that is determined by the
$D$-brane geometry. The space $V$ may be thought of as the
appropriate concept for systematically combining
differential forms on the \CY $X$ (describing the closed
string sector) with forms on the boundary $\Z$ (describing 
the open string sector). 

The relative cohomology group $V$ comes with an important
mathematical package, namely the mixed Hodge structure,
which defines an integral grading, $V=\oplus_q V^{(q)}$, on it.
Moreover, the Gauss-Manin connection $\nabla$ on the relative
cohomology bundle over $\MM$ (with fibers $V$) represents
an integrable connection on the space of topological 
B-models. The integrability implies the existence of
special, topological coordinates $t_A$ on $\MM$,  
in which the Gauss-Manin derivatives reduce to ordinary derivatives.

Moreover, the geometric representation of the open-closed chiral ring
allows to describe its moduli dependence 
by the ``relative period matrix'' $\Pi_I^\Si$.
It consists of the period integrals of the elements of $V$,
with respect to a basis of topological cycles with boundaries on $\Z$.
In the special coordinates $t_A$,
the relative period matrix takes the form:
\eqn\pmii{
\Pi^\Si_I \ 
=\ \pmatrix{
1&t_A&\cx W_K&...\cr
0&\delta_{AB}&\p_B\cx W_K&...\cr
0&0          &C_{AB}^{\ \ K}&...\cr
\vdots &\vdots &\vdots\cr}
.}

\noi
The first row, $\Pi^\Si_0$, coincides
with the relative period vector $\Pi^\Si$ which enters the definition
of the superpotential $W$ in \gensp. 
The entries $t_A(z_\dd)$ of the relative period matrix define
the {\it $\cx N=1$ mirror map}, in terms of the ratios of certain
period and chain integrals on $X$. This map determines 
the special coordinates $t_A$ in terms of arbitrary 
coordinates $z_A$ on $\MM$.

The entries $\cx W_K$ of the relative period matrix
represent the {\it holomorphic potentials of $\cx N=1$ special
geometry}. They are, in a quite precise sense, 
the $\cx N=1$ counterparts of the holomorphic 
prepotential $\cx F$ of $\cx N=2$ special geometry.\foot{
Note that a subset of the $\cx W_K$ may coincide with the
derivatives, $\cF_a$, of the bulk prepotential. However, in general
the $\cx W_K$ cannot be integrated to a single prepotential;
the fact that one needs more than one holomorphic function to define the $\cx N=1$ special geometry, reflects the fact that it is less restricted than 
the special geometry associated with $\cx N=2$ space-time 
supersymmetry.}
In particular, their derivatives determine the structure constants of the 
2d chiral ring \rocky:
\eqn\ocr{
C_{AB}^{\ \ K}(t_\xd)=\p_A\p_B \cx W_K(t_\xd).}
Moreover, the potentials $\cx W_K$ have also an important 
physical meaning as they encode the space-time
instanton expansion of the $\cx N=1$ superpotential 
in the string effective space-time theory. This
expansion has been conjectured to have 
remarkable integrality properties \OV. Specifically,
when expressed in terms of the special coordinates $t_A$, 
the potentials have the following expansions:%
\eqn\Winstexpan{
\cx W_K(t_\xd)\ =\
\sum_{\{n_C\}} N^{(K)}_{n_1\dots n_M}
\sum_k\fc{1}{k^2}\big(\prod_C e^{2\pi i k n_Ct_C}\big),
} 
where the coefficients $N^{(K)}_{n_1\dots n_M}$ are supposedly integral.
In writing the above expression, we have assumed that the physically
motivated definition of the special coordinates given in \doubref\OV\AVi, 
namely in terms of the tension of certain domain walls, 
is equivalent to the above 2d definition in terms of the vanishing of the 
Gauss-Manin connection. This is a quite non-trivial and important
relation. The results in the following sections give substantial 
evidence for the equivalence of these two definitions.

Note that the coefficients $N^{(K)}_{n_1\dots n_M}$ have, besides 
other interpretations, a well-known interpretation in the language of the A-model mirror as corrections from
world-sheet instantons of disc \doubref\OV\AVi\ and sphere \doubref\PMsusy\CVlargeN\ topologies. However, since the scalar vev's $t_A$ determine the coupling constants of the RR
sector in the four-dimensional string effective supergravity,  the
expression \Winstexpan\ can often be directly interpreted, alternatively, also as corrections to the $\cx N=1$ superpotential from space-time instantons.

In the following, we will work out the details of
the geometrical structure outlined above;
the next two sections provide a general
discussion of the topological field theoretic and geometric
aspects, while in sects.~5 and 6 we will present some 
additional structure that is specific to toric geometries, 
and some explicit computations in such geometries.

\goodbreak

%%%%%%%%%%%%%%%%%%%%%%%%%%%%%%%%%%%%%%%%%%%%%%%%%
\newsec{Observables in the open-closed B-model and relative cohomology}
%%%%%%%%%%%%%%%%%%%%%%%%%%%%%%%%%%%%%%%%%%%%%%%%%

As is well-known, the elements of the chiral ring in the 
closed string B-model on the \CY $X$ have an geometric
representation as elements of the cohomology group $H^3(X)$ \wittop.
Moreover the gradation by $U(1)$ charge of the chiral ring
corresponds to the Hodge decomposition $H^3(X)=\oplus_q H^{3-q,q}(X)$.
We will now describe a similar representation of the chiral ring
$\Roc$ of the open-closed B-model in terms of a certain
relative cohomology group.

The open-closed chiral ring is an extension of the 
closed string chiral ring, and combines operators
from both the bulk and boundary sectors.
Geometrically, the new structure from the open string
sector is the submanifold $B\subset X$ wrapped 
by the D-branes.\foot{We will restrict our discussion to
two-cycles and trivial line bundles on them.} 
Since the bulk sector of
the closed string is represented by $H^3(X)$, the
open-closed chiral ring should correspond to an
extension of this group by new elements originating
in the open string sector on $B$.
It is natural to expect that this extension is 
simply the dual $H^3(X,B)$ of the space $H_3(X,B)$ that
underlies the flux and brane induced superpotential \gensp.

Under a certain assumption discussed in \Let, 
one may in fact replace the group $H^3(X,B)$ 
by a simpler relative cohomology group, $H^3(X,\Z)$. 
Here $\Z$
denotes a union of hypersurfaces that pass through the 2-cycles in $B$ 
wrapped by the 5-branes.
Let us here recall briefly the geometry of the hypersurface $\Z$
and the assumption which underlies it.
As in \Let, we will restrict the discussion to 
a single 2-cycle $B$ with a single modulus. In fact 
the superpotential for a collection of 
non-intersecting branes wrapped on a set of 2-cycles $\{B_\nu\}$
will be the sum of the individual superpotentials and may be treated
similarly.

One way to solve the minimal volume condition for 
a special Lagrangian 3-cycle
$\Ga$, is to slice it into a family of 2-cycles along a path $I$.
This can be achieved by intersecting $\Ga$ with a suitable
one-parameter family of holomorphic hypersurfaces $\Z(\z)$, 
with $\z$ a complex parameter. The intersection
of the hypersurface $\Z(\z)$ with $\Gamma$ is a family of 2-cycles
$\Gamma_2(\z)$ of minimal volume $V(\z)$ and 
the integral \lpv\ can be written as
\eqn\wgamma{
W_\Gamma= \int_\Gamma \Om = \int_{\z_0}^{\z_1}V(\z)d\z.
}
Here the contour $I$ in the $\z$-plane is determined by the
minimal volume condition for $\Gamma$. 

For a 3-chain $\hx\Gamma$,
the interval in the $\z$-plane ends at a
specific value, say $\z=\hx z$, for which
the hypersurface $\Z(\hx z)$ passes through the 
boundary 2-cycle $\C_\nu=\p\hx\Gamma^\nu\subset \Y$ 
(around which the 5-brane is wrapped).
Varying the position of the 5-brane leads to the following variation 
of the brane superpotential:
$$
\delta W_{\hx\Gamma}\ \sim\  V(\hx z)\ =\ 
\int_{\Gamma_2(\hx z)} \om,
$$
where $\om$ is an appropriate holomorphic form on $\Z(\z)$. 
As will be discussed below, 
$\om$ is the $(2,0)$ form on $\Z(\z)$ obtained from a
Poincar\'e residue of $\Om$. 
Note that $\delta W_\Gamma$ vanishes if $\Gamma_2$ is holomorphic 
\refs{\WQCD,\AVi,\KKLM}.

The moral of the foregoing discussion is that the variations of the 
relative period vector \lpv, related to the boundary of a chain
$\Gamma$, are detected by the periods of 
a holomorphic $(2,0)$ form $\om$ on the hypersurface $\Z(\z)$.\foot{
See also \Grifdeq\ for a related discussion.}
In other words, to describe the variations of the volumes of the special
Lagrangian chains, we may replace the relative cohomology group 
$H^3(X,\Y)$ by the group $H^3(X,\Z)$. This connection will 
be made explicit in the next section, by associating the topological
observables of the open-closed B-model with elements of the
relative cohomology group $H^3(X,\Z)$.

\subsec{Space of RR ground states and relative cohomology}

The closed string B-model for the \CY 3-fold $X$ 
is defined by a twisted, two-dimensional $N=(2,2)$ superconformal
theory on the string world-sheet $\Sigma$ \wittop. 
The map $\vphi:\, \Sigma \to X$ from the
world-sheet to the target space $X$ is defined by three complex 
scalar fields $\vphi^i$ on $\Sigma$. The fermionic superpartners
are three complex fermions $\psi^i_\pm$ which are sections of
the holomorphic tangent bundle $T=T^{1,0}(X)$. 
The BRST cohomology is generated by the linear 
combinations
\eqn\tdfs{\eqalign{
\eta^{\bb i}&=\psi_+^{\bb i}+\psi^{\bb i}_- \,,\cr
\theta_i     &=g_{i\bb j}(\psi_+^{\bb j}-\psi^{\bb j}_-)\,,
}}
which satisfy $\delta_Q \eta=\delta_Q \theta = 0$;
here $Q$ is the BRST operator which corresponds to $\bb \p$.
The local BRST observables have the form 
\eqn\brstops{
\phi_{\ib_1,...,\ib_p}^{j_1,...,j_q}
\eta^{\ib_1}...\eta^{\ib_p}\theta_{j_1}...\theta_{j_q}.
}
The local BRST cohomology is thus isomorphic to the 
space of 
sections of $H^p(X,\wedge^q T)$.

In the open string sector, the B-type boundary conditions 
set to zero a certain linear combination
of the fermionic fields, in the simplest form one has \doubref\Detal\HIV:
\eqn\bcs{\eqalign{
\eta^{\bb i}&=0\ \ \ (D),\cr
\theta_i&=0\ \ \ (N),
}}
where $D$ and $N$ denote a Dirichlet and Neumann boundary condition
in the $i$-th direction, respectively. 
The cohomology of the boundary BRST operator 
can also be represented by elements of the form \brstops,
however with the interpretation as sections of 
$H^p(\Z,\wedge^q N_\Z)$.
Here $N_\Z$ denotes the normal bundle to the submanifold $\Z$ 
in $X$
on which the Dirichlet boundary conditions are imposed.
More
generally, one may couple the fermions to a gauge bundle $E$ on the
D-brane, leading to a generalized BRST operator $\bb \p_{E}$
acting on a Hilbert space isomorphic to the space of 
sections $H^p(\Z,\wedge^q N_\Z\otimes E)$ \WitCS.

We interpret a section $\theta \in H^0(\Z,N_\Z)$ as the 
restriction $\theta=\Theta|_\Z$ of a $C^\infty$ section $\Theta$
of some bundle $E$ on $X$. The zero locus of the section 
$\Theta$ defines a submanifold in $X$ that can 
be wrapped by a D-brane. Deformations of the section $\Theta$, 
corresponding to a variation of the location of the D-brane, represent 
the open string moduli $\th_\al$. These deformations correspond to 
physical fields, $\ph_\al$, of $U(1)$ charge one\foot{Quantities
in the open string sector 
will be denoted with a hat and greek indices;
letters from the beginning of the alphabet refer to fields with 
$U(1)$ charge one.}, associated to the section $\theta_\al$:
\eqn\ospf{
\ph_\al \ \longleftrightarrow\  \theta_\al \in  H^0(\Z,N_\Z).}
On the world-sheet the deformations correspond to adding 
boundary terms 
$\delta S = \sum_\al \th_\al \int_{\p \Sigma} dz d\theta \hx \Phi_\al$.
See \HMA\ for a discussion of ordering effects involving
such terms.

There are two basic multiplications
induced by wedge products with sections of $H^0(\Z,N_\Z)$:
%\vskip-9pt
\eqn\ecr{\eqalign{
\ph_{\al}\cdot \ph_{\be}&=
C_{\al\be}^{\ \mu}\ph_{\mu}
\,\ \longleftrightarrow\  \,
H^0(\Z,N_\Z) \times H^0(\Z,N_\Z) \to H^{0}(\Z,\wedge^{2} N_\Z),\cr
\phi_a \cdot \ph_\al&= C_{a\al}^{\ \mu}\ph_\mu
\, \ \longleftrightarrow\ \,
H^1(X,T\ )\, \times H^0(\Z,N_\Z)\to H^{1}(\Z,T|_\Z\wedge N_\Z ).
}}
Together with the closed string chiral ring operator product \ccr,
these products generate generate the extended, open-closed chiral ring $\Roc$.
Here it is understood that the open and closed string operators 
are inserted on a world-sheet with boundary, and the restriction of 
the closed string operator to the boundary is defined by the geometric
restrictions $T\to T_{X|\Z}$ and $T^*~\to~T^*_{X|\Z}$. 

More judiciously one should replace the bulk operators $\phi_a$ in
\ecr\ by a set of boundary operators $\ph_a$ defined
by the collision of the bulk operators $\phi_a$ 
with the boundary. At the level of the topological observables 
considered in this paper, this step is already built into the
geometric representation $H^3(X,\CC)$ of the ``closed''
string Hilbert space by the following simple observation.
The integral dual homology cycles in $H_3(X,\ZZ)$ define a basis of D-branes
wrapped on the 3-cycles of $X$. The transition
from bulk to boundary observables is the obvious pairing 
$H_3(X)\times H^3(X) \to \CC$ \doubref\OOY\HIV. The
bulk-boundary product is thus implicit in the
following discussion when passing to a flat basis dual
to $H_3(X,\ZZ)$.

Similarly as was done in \BCOV\ for the closed B-model, 
we would like to describe
the space spanned by the elements of the chiral ring by
a geometric cohomology group $V$. Recall that the chiral fields $\phi_I$
in the NSNS are in one-to-one correspondence with the 
supersymmetric ground states in the RR sector. In particular a
ground state $\bra I$ can be obtained from the canonical 
vacuum $\bra 0$ by inserting the operator $\phi_I\in \{\phi_a,\ph_\mu\}$ 
in the twisted 2d path integral on a world-sheet
with boundary, $\bra I = \phi_I\, \bra 0$ \BCOV. The 
cohomology group $V$ may thus be alternatively identified with the space of 
2d RR ground states.

As is well-known, the elements of the bulk chiral ring arising from
sections of $H^p(\wedge^p T)$ can be identified 
with sections of $H^{3-p,p}(X)$ by the isomorphisms
\def\rhom{\rho_{\, \Om}}
\eqn\defrho{
\rhom: \ H^p(\wedge^p T) \lra{\Om} H^{3-p,p}(X,\CC)\,,
}
provided by the unique holomorphic $(3,0)$ form $\Om$. Importantly, the
holomorphic $(3,0)$ form defines also an 
isomorphism on the elements in the open string sector 
associated with the boundary $\Z$,
\eqn\defrhoii{
\rhom:\ H^0(\Z,N_\Z)\lra{\Om} H^{2,0}(\Z),\quad
       H^1(\Z,T|_\Z\wedge N_\Z)\lra{\Om} H^{1,1}(\Z)
}
This isomorphism can be interpreted as a Poincar\'e residue for $\Om$,
which may be written as \Griii
\eqn\defpr{
\tx \phi(s_1,...,s_p;\theta_\al)= \phi|_\Z
(s_1,...,s_p,\Theta_\al), \qquad s_k \in T_\Z
}
where $\tx \phi\in H^{p,0}(N^*_\Z)$  and $\phi$ is
a holomorphic $(p+1)$-form on $X$.\foot{This definition, which
is directly related to \defrhoii,  is
equivalent to the more familiar definition used e.g. in \Can.} 
%Here $\Theta_\al|_\Z=\theta_\al \in H^0(\Z,N_\Z)$
%is the section associated to the open string operator $\ph_\al$.

For closed strings, the 
isomorphism $\rhom$ has the simple interpretation of a choice of 
normalization for the canonical ground state $\bra 0\sim H^{3,0}$ \BCOV.
In the next section we define a grading of the space $V$ of open string
RR ground states that again identifies
$H^{3,0}$ with the canonical vacuum. Thus
$\Om$ has precisely the same meaning also in the open string context, with
the space $H^{3,0}$ identified as the RR vacuum 
$\bra 0$ on world-sheets with boundaries, 
on which the operators $\phi_i$ and $\ph_\mu$ act to
generate the full space $V$ of RR ground states.

We are now ready to argue that 
the total space of RR vacua can be described 
by the relative cohomology group $H^3(X,\Z)$. The
latter is defined as follows. The inclusion $i:\Z\to
X$ induces the morphism of the complexes of sheaves
%of holomorphic forms on $X$ and on $\Z$
$
i^*:\, \Om_X^* \to  \Om_\Z^*.
$
The mapping cone of $i^*$ is defined as the complex 
$\Om_X^*\oplus \Om_\Z^{*-1}$ with differential 
$d=(d_X+i^*,-d_\Z)$. The relative cohomology is the 
hyper cohomology of this complex. From the exact sequence: 
$$
0\longrightarrow{} \Om^n(X,\Z)\longrightarrow{}\Om^n(X)\longrightarrow\Om^{n}(\Z)\longrightarrow{}0
$$
it follows that the $n$-th hyper cohomology is the space
$$
\rmx{Ker}[H^n(X)\to H^n(\Z)]\oplus \rmx{Coker}[H^{n-1}(X)\to H^{n-1}(\Z)].
$$
In the present context, with $V=H^3(X,\Z)$, 
the first factor will be all of $H^3(X)$ for the D-brane geometries that
we will consider. This statement will be
easy to check in practice, as $H^3(\Z)$ is trivial for these
geometries.
The second factor is the
variable cohomology\foot{The variable cohomology
is the part of the cohomology which varies with the embedding $i:\, \Z\to X$.
E.g. the cohomology of the quintic $\Z$ in $\IP^4$ splits over the rationals 
into the 
even-dimensional cohomology $H^{2n}(\Z)=i^*H^{2n}(\IP^4)$, 
which is the ``fixed''
cohomology inherited from the ambient $\IP^4$, and
the odd-dimensional ``variable'' cohomology $H^3(\Z)$ which does not descend
from the ambient space.} $H^2_{var}(\Z)$ of 
$\Z$.
The image of the Poincar\'e residue precisely maps into this space.
In summary we thus have\foot{In many cases the following inequality will be
an equality; however this statement may sometimes fail for a 
chosen algebraic realization of the geometry.}

\eqn\rrgsts{
\rmx{Space\ of\ RR\ ground states}\ V \hskip 4pt \sim\hskip 4pt 
\rmx{Im}(\rhom)\ \subseteq\ H^3(X,\Z)\ .
}

%%%%%%%%%%%%%%%%%%%%%%%%%%%%%%%%%%%%%%%%%%%%%%%%%
\newsec{Variations of mixed Hodge structures and topological flat 
connection}
%%%%%%%%%%%%%%%%%%%%%%%%%%%%%%%%%%%%%%%%%%%%%%%%%

The next step is to extend the previous discussion to families of the
open-closed B-model over the space $\MM$ of open-closed string
deformations. The main concepts here will be the variation of 
mixed Hodge structures \MHV\ defined on the relative cohomology group
$H^3(X,\Z)$, and a topological flat connection $\nabla$ on the
vacuum bundle $\cx V$ with fibers $V\sim H^3(X,\Z)$.
The connection $\nabla$ is the Gauss-Manin connection
on the relative cohomology bundle with fiber $H^3(X,\Z)$. 

\subsec{Infinitesimal deformations and variations of mixed Hodge structure}

Our first aim is to identify geometric cohomology classes that are related to
infinitesimal variations in the moduli and generate the 
space $V=H^3(X,\Z)$ of RR ground states from a ``canonical vacuum''.
This leads to a natural 
gradation 
\eqn\vgrad{
V=\oplus_qV^{(q)},}
which can be identified with 
a $U(1)$ charge in the TFT. In the
present case it will also be necessary to distinguish between
infinitesimal variations in the bulk and boundary sectors, respectively.
The resulting filtration is in fact 
well-known in mathematics and defines a mixed Hodge structure on 
the relative cohomology group $V$.

In relative cohomology\foot{See e.g., \Kar\ for an introduction.}, 
a RR ground state may be represented  by a pair of forms 
\eqn\Thetadef{
\vx \Theta\ =\ (\Theta_X,\theta_\Z)\ \in\ H^3(X,\Z),
}
where $\Theta_X$ is a closed 3-form on $X$ and $\theta_\Z$ a 2-form on $\Z$ 
such that $i^*\Theta_X-d\theta_\Z=0$. The equivalence relation is
\eqn\relc{
\vx \Theta \sim \vx \Theta + (d\om,i^*\om-d\phi)\ ,
}
with $\om$ ($\phi$) a 2-form on $X$ (1-form on $\Z$).

We will start from the ansatz
\eqn\defcom{ (\Om,0)}
as a representative for the canonical ground state $\rbra 0$.
To identify the geometric classes that generate
the total space $V$ from $\rbra 0$, 
consider the infinitesimal deformations for a family of
B-models defined by the geometry $(X,\Z)$. 
The moduli space $\MM$ of open-closed string deformations
of the geometry $(X,\Z)$ consists of complex deformations
of $X$ and deformations of the sub-manifold $\Z\subset X$. 

The general geometric argument which identifies 
infinitesimal variations in the moduli with geometric classes 
goes as follows. Consider an analytic family  
$\cx A =\{A_b\}_{b\in \Delta}$, where $\Delta$ is a disc that 
represents a local patch in the moduli space for an algebraic 
variety $A_b$. 
The infinitesimal displacement mapping of Kodaira is
\eqn\kodm{
T_b(\Delta)\longrightarrow H^0(A_b,N_{A_b}),}
where $N_{A_b}$ is the normal bundle of $A_b$ within $\cx A$. 
On the other hand, the co-boundary of the exact sequence
\eqn\tbseq{
0\longrightarrow T_{A_b} \longrightarrow T_{\cx A|A_b} \longrightarrow
\pi^*T_\Delta|_{A_b} \longrightarrow 0, }
where $\pi:\, \cx A \to \Delta$, gives the Kodaira-Spencer map:
\eqn\kodspm{
\kappa:\ T_b(\Delta) \longrightarrow H^1(T_{A_b}).}
Applying the above maps to the present geometric situation,
one obtains altogether four classes, which 
define the action of the 
closed and open string deformations, respectively, on
$\Theta_X$ and $\theta_\Z$ in \Thetadef. 

Explicitly, this works as follows.
Consider first the family of \CY manifolds $\cx X \to \cx M_{CS}$
with fiber $X_{\{t_a\}}$. The Kodaira-Spencer map for this family gives
a class in $H^1(X,T_X)$ that defines a map on the first 
entry of $\vx \Theta$ by
taking the wedge product and contracting the form with the vector field.
This map represents
the geometric counter-part of the well-known TFT argument
that identifies an infinitesimal variation $\delta_a$ in the 
closed string moduli with
multiplication by a charge one operator $\phi_a \in H^1(X,T_X)$.
Restricting the sequence \tbseq\ to $\Z$ one obtains
a class in $H^1(\Z,T_X|_\Z)$ 
that acts on the second entry of $\vx \Theta$ in a similar way.

As for an open string variation $\hx \delta_\al$, one may likewise consider
the Kodaira-Spencer map for the family $X\to \Delta_z$ with fiber $\Z(z)$ to
obtain a class in $H^1(\Z,T_\Z)$ that defines a map 
on the second entry of $\vx \Theta$.
On the other hand, the action of $\hx \delta_\al$ on the first entry 
of $\vx \Theta$, 
which is a form on $X$, cannot be obtained from the Kodaira-Spencer map;
one really needs a class on the total space $X$ of the family.
Such a class may be defined by the Kodaira map \kodm, 
and it acts on the first entry by the Poincar\'e residue \defpr.

The common property of the maps defined by these 
four classes associated with the open-closed
string deformations, is that
they all lower the holomorphic degree of a form by one.
To complete the discussion of how these classes generate the 
space $V$ and define a natural gradation on it, let us pass right away
to the mathematical definition of a mixed Hodge structure \MHV\
on $V$ which is the appropriate language for the present problem. 

A mixed Hodge structure is
defined by $i)$ a $\ZZ$ module $V_\ZZ$ of finite rank;
$ii)$ a finite decreasing Hodge filtration $F^p$ on
$V=V_\ZZ\otimes \CC$; $iii)$ a finite increasing weight
filtration $W_p$ on $V_{\bx Q}=V_\ZZ\otimes Q$ such that
$F^p$ defines a pure Hodge structure of weight $p$
on the quotient $Gr^W_p=W_p/W_{p-1}$.
In the present case we have $V=H^3(X,\Z)$ and the filtrations\foot{In the 
following equation we have used 
the Poincar\'e residue to write a basis of the relative cohomology for 
$(X,\Z)$ in terms of elements of the cohomologies on $X$ and $\Z$. 
This will be also used below. More generally,
the filtrations are defined on the hypercohomology
of the complexes $\Om^*_X(\rmx{log}\, \Z)$. }
are

\vbox{
\vskip10pt
\eqn\filts{
\vbox{\offinterlineskip\tabskip=0pt\halign{
\strut
$#$\hfil
&\hskip 66pt $#$\hfil\cr
F^3:\ H_X^{3,0},
&W_3:\ \oplus_{p+q=3} H_X^{p,q}\cr
F^2:\ F^3+ H_X^{2,1}+H_\Z^{2,0}
&W_4:\ W_3+\oplus_{p+q=2}H_\Z^{p,q}\cr
F^1:\ F^2+H_X^{1,2}+H_\Z^{1,1},&\cr
F^0:\ F^1+H_X^{0,3}+H_\Z^{0,2},&\cr
}}}}
\noi
The quotients $Gr^W_p$ separate
the observables into bulk and boundary operators on 
which separate Hodge filtrations may be defined. 
On the other hand, the quotients $Gr^F_p=F^p/F^{p+1}$ define 
the spaces \vgrad\ of pure (left) $U(1)$ charge in the TFT. The motivation
to prefer the filtered spaces $F^p$ over the
spaces of pure grades defined by the quotient is the same 
as in the closed string case: the bundles over $\MM$ with fibers $F^p$
are holomorphic bundles, whereas the bundles with the quotient space
as the fibers are not.

In terms of the above basis, the action of the infinitesimal 
variations on $V$ can be written more explicitly as 
\eqn\ocvi{\eqalign{
\delta_a \in\ \  &
\Hom(H^{p,q}(X),H^{p-1,q+1}(X))\oplus \Hom(H^{p,q-1}(\Z),H^{p-1,q}(\Z))\cr
\hx \delta_\al\in\ \  &
\Hom(H^{p,q}(X),H^{p-1,q}(\Z))\hskip 12pt 
\oplus \Hom(H^{p,q-1}(\Z),H^{p-1,q}(\Z)).\cr
}}
Schematically, the open-closed string deformations act as
%ddd
\eqn\ocdia{\xymatrix{
(3,0)_X\ar[r]^{\ss \delta_a}\ar[dr]^{\ss \delh_\al}&
(2,1)_X\ar[r]^{\ss \delta_a}\ar[dr]^{\ss \delh_\al}&
(1,2)_X\ar[r]^{\ss \delta_a}\ar[dr]^{\ss \delh_\al}&
(0,3)_X\ar[dr]^{\ss \delta_a,\delh_\al} \\
&(2,0)_\Z\ar[r]^{\ss \delta_a,\delh_\al}&
(1,1)_\Z\ar[r]^{\ss \delta_a,\delh_\al}&
(0,2)_\Z\ar[r]^{\ss \delta_a,\delh_\al}& 0}}
{}From the maps defined in the above diagram one may
obtain an explicit basis for the extended ring $\Roc$ at each point
in the moduli space $\MM$, given 
the classes $\delta_a$  and $\delh_\al$ at that point.\foot{For 
intersecting brane configurations, there 
may be additional steps in the vertical direction in \ocdia, describing
cohomology at the intersections of higher codimension.} 
The appropriate framework to address the latter
question is to construct the topological connection $\nabla$ on the 
family of B-models reached by finite deformations, which 
will be discussed now.

\subsec{Moduli dependent open-closed chiral ring 
and topological flat connection}

The spaces $V=H^3(X,\Z)$ for a family of open-closed B-models
parametrized by some coordinates $z_A\in \MM$ 
are all isomorphic. 
The bundle $\cx V$ over $\MM$, with fiber $V(z_A,\bb z _{\bb A})$, 
is thus locally constant and admits a flat
connection $\nabla$. This is  the 
Gauss-Manin connection on the relative cohomology
bundle with fiber $H^3(X,\Z;\CC)$.
The existence of a flat connection on the family of twisted, 
two-dimensional $\cx N=2$ SCFT's on the world-sheet has been
predicted in full generality in \CeVa\ as a consequence
of the so-called $tt^*$ equations.
Their arguments also apply to world-sheets
with boundaries, although the content of the $tt^*$ equations
for the open-closed B-model
has not been worked out so far\foot{For
a discussion of some aspects of the open string case, see 
\BCOV.}. We proceed with the 
geometric approach provided by the open-closed B-model; clearly it
would be interesting to study and extend the results
in the full framework of $tt^*$ geometry.

To complete the geometric representation of the open-closed
chiral ring, we are looking for: 
$i)$ A graded basis of cohomology classes $\Phiv q _I(z_A)$ for $\cx V$, that 
represents the elements of the chiral ring $\Roc$; $ii)$ Flat
topological coordinates $t_A(z_B)$ 
on the moduli space $\MM$ that correspond
to the coupling constants in the 2d world-sheet Lagrangian.
These are, on very general grounds \BCOV, the good coordinates for
the instanton expansion of the correlation functions.
The topological basis $\{\Phiv q _I (t_A)\}$ is singled out by the property
\eqn\ringrel{
\fc{\p}{\p t_A} \Phiv q _I(t_B) = C_{AI}^{\ K}(t_B)\, \Phiv{q+1} _K(t_B).
}
This equation expresses that an infinitesimal variation in the
$t_A$ direction is equivalent to multiplication by the grade 
one field $\Phiv 1 _A(t_B)$ in the chiral ring. Moreover 
the basis $\{\Phiv q _I (t_A)\}$
is generated from the canonical vacuum by repeated application of 
\ringrel.

To guarantee holomorphicity of the bundle, of which the $\Phiv q _I(z_A)$ are
sections\foot{We continue to 
denote arbitrary coordinates on $\MM$ by $z_a$ and the flat topological
coordinates by $t_A$.},
one must define the classes $\Phiv q _I (z_A)$ as an element
of the Hodge filtration $F^{3-q}$ rather than of pure grade $q$. In
particular the bundles $\cx F^{q}=F^{q}\cx V$ are holomorphic
sub-bundles of $\cx V$, whereas the bundles with fibers $Gr_F^q\cx V$
are not holomorphic. In this way one may 
preserve holomorphicity of the basis for the vacuum bundle;
one may always project the $\Phiv q _I(z_A)$ to the 
pieces of highest grade $3-q$ at the end to obtain eventually a basis of pure
Hodge type (varying non-holomorphically with the moduli $z_A$).

To construct the basis $\{\Phiv q _I \}$, we start from the unique element
$\Phiv 0 (z_A) = (\Om(z_A),0) \in \cx F^{3}$. 
The multi-derivatives $\p/\p z_{A_1}...\p/\p z_{A_k}\,\Om$
for $k\leq 3$ generate a basis $\{\Psiv q _I\}$ of $V$, where 
$\Psiv q _I(z_A)\in \cx F^{3-q}$. 
This is a consequence of Griffiths' transversality%
$$
\nabla: \ F^p\cx V \lra{} F^{p-1}\cx V \otimes \Om_{\MM},
$$%
which says that the action of the Gauss-Manin derivative increases 
the grade of the Hodge filtration by one. 
Note that the derivatives $\p/\p z_A$
differ from the covariant derivatives $\nabla_A$ by terms which 
do not increase the grade and thus are irrelevant at this step. 
Note also that in general not all of the multi-derivatives 
$\p/\p z_{A_1}...\p/\p z_{A_l}\, \Om$ for $l\leq k$ 
will be linearly independent for given $k$,
but only $\rmx{dim}(F^{3-k}V)$ of them. 

One may describe the moduli 
dependence of the elements $\Psiv q _I(z_A)$ by expressing them in terms
of a flat, moduli independent basis $\{\vx \Ga_\Si\}$ for $\cx V$:
\eqn\basistf{
\Psiv q _I(z_A) = \Pi_I^\Si(z_A) \, \vx \Ga_\Si.
}
The basis $\{\vx \Ga_\Si\}$ may be defined as the basis dual to a
constant basis $\{\Ga^\Si\} $ of moduli independent cycles  in $H_3(X,\Z)$.
The transition matrix 
\eqn\relperma{
\Pi_I^\Si(t)\ = \ \langle \Ga^\Si, \vx \Psi_I\rangle,}
is the {\it relative period matrix}. The pairing 
$\langle \Ga^\Si,\vx \Theta \rangle$ in relative co-homology  is defined by 
\eqn\relpair{
\langle \Ga^\Si, \vx \Theta\rangle\ = \ 
\int_{\Ga^\Si} \Theta - \int_{\p \Gamma^\Si}\theta,
}
for $\vx \Theta = (\Theta,\theta)\in H^3(X,\Z)$ and 
$\Ga^\Si \in H_3(X,\Z)$. It is the appropriate 
pairing in relative co-homology, $H^3(X,\Z)\times H_3(X,\Z)\to \CC$, 
invariant under the 
equivalence relation \relc.

Let us order the basis $\{\Psiv q _I \}$ by increasing grade $q$.
The searched for basis $\{\Phiv q _I \}$, representing the chiral
ring,  may be obtained from the basis $\{\Psiv q _I \}$ by a 
linear transformation with  holomorphic coefficient functions
that preserves the Hodge filtration. By such a 
transformation one may put  the relative period matrix $\Pi_I^\Si$ 
into upper triangular form, implying the relations 
\eqn\betterbasis{
\nabla_A \Psiv q _I(z_B) = C_{AI}^{\ K}(z_B)\,  \Psiv{q+1}_K(z_B),
}
where we use again $\Psiv q_I (z_A)$ to denote the elements of
the transformed basis. It follows that the piece of pure type 
$d-q$ in $\Psiv{q+1} _I$ represents the chiral ring, up to
a moduli dependent normalization. 

One may further find the {\it flat} topological coordinates 
$t_A(z_B)$ by requiring the connection pieces in $\nabla_A$ to
vanish. This is achieved by a moduli dependent, holomorphic 
change of normalization of the classes $\Psiv q _I $ that makes
the block-diagonal 
terms in the transition matrix $\Pi_I^\Si$ constant
in this basis\foot{In particular the diagonal block at grade $q=1$
can be chosen to be the unit matrix {\bf 1}$_M$.}. 
The classes so obtained provide the searched
for basis $\{\Phiv q _I\}$ that represents the open-closed 
chiral ring $\Roc$. 

Note that the flat coordinates $t_A$, as defined by the above 
procedure, are the solutions of 
$$
\nabla_A \Pi^{\Si}= \p_A \Pi^{\Si}= \delta_{A}^{\Si}, \qquad
\Si'=0,...,M,\ A=1,...,M.
$$
Here $\Pi^\Si\equiv\Pi^\Si_0$ is the relative period vector representing 
the first row of the period matrix and $M=\rmx{dim}\, \MM$.
Thus the flat coordinates $t_A$ are given by 
ratios of relative period integrals, and this is what determines
the $\cx N=1$ mirror map $t_A(z_B)$ \PM:
\eqn\flatco{
t_A(z_B)=\fc{\langle \Ga_A, \Phiv 0 (z_B)
 \rangle}{\langle \Ga_0, \Phiv 0 (z_B)\rangle},
\qquad A=1,...,M.
}
For the choice of vacuum $\Phiv 0 = (\Om,0)$, this expression is 
identical in form to the well-known mirror map for the flat 
coordinates for the closed string with 
$\cx N=2$ supersymmetry \doubref\Can\GMP.
However, in the present context, it describes the flat 
coordinates on the space of chiral $\cx N=1$ multiplets
$\MM$, with the pairing in \flatco\ defined in
relative co-homology.
Another definition of flat coordinates had been 
previously given in \refs{\OV,\AVi,\AVii}, in terms
of the tension of D-brane domain walls; it
leads to the same functional dependence $t_A(z_A)$,
at least for the class of open-closed compactifications
studied so far. 

Finally note that the equations \betterbasis\ 
determine the chiral ring coefficients 
for grade $q>1$ in terms of the 
derivatives of the relative period matrix.
In fact, it follows from eq.\betterbasis\ that the relative period 
matrix $\Pi_I^\Si(z_A)$ satisfies the system of linear differential equations

\eqn\pfoc{
(\nabla_A - C_A)\, \Pi_I^\Si(z_A) = 0.}
Eliminating the lower rows of $\Pi_I^\Si$ in favor of the first row,
one may thus obtain all chiral ring coefficients $C_{AI}^{\ K}$ 
from derivatives of the relative period vector $\Pi^\Si$. In particular, 
at $q=2$ one obtains the promised relations
\eqn\holpots{
C_{AI}^{\ K} = \p_A\p_B \cx W_K,
\qquad \cx W_K = \langle \Ga^\Si, \Phiv 2 _K
\rangle,}
which express the holomorphic potentials $\cx W_K$ of $\cx N=1$
special geometry as certain entries of the relative period matrix.
In the flat coordinates $t_A$, the 
first rows and columns of $\Pi_I^\Si$ take the form

\eqn\pmiii{
\Pi^\Si_I = \langle \Ga_\Si,\, \vx \Phi_I \rangle 
=\pmatrix{
1&t_A&\cx W_K&...\cr
0&\delta_{AB}&\p_B\cx W_K&...\cr}.}

\vskip5pt
By iterative elimination of the lower rows in \pfoc\ one obtains
a system of homogeneous, linear differential equations for 
the period vector $\Pi^\Si$ of higher order. This is 
a Picard-Fuchs system
for the relative cohomology group $H^3(X,\Z)$. One may
determine the exact expressions for the mirror map 
\flatco\ and the holomorphic potentials $\cx W_K$ 
from the solutions to the Picard-Fuchs system with
appropriate boundary conditions by standard methods
(see \doubref\PM\LM\ for some explicit examples).

As it turns out,  this approach is particularly powerful 
for $D$-branes on \CY manifolds that can be represented as hypersurfaces in toric varieties; this class of geometries will be discussed in the next section.

%%%%%%%%%%%%%%%%%%%%%%%%%%%%%%%%%%%%%%%%%%%%%%%%%
\newsec{Relative cohomology in toric geometry, and GKZ type differential equations}
%%%%%%%%%%%%%%%%%%%%%%%%%%%%%%%%%%%%%%%%%%%%%%%%%

In this section we illustrate the previous ideas in detail for 
a class of open-closed string \CY backgrounds that may be 
realized by a linear sigma model (LSM) \Witlsm. The 
aim is to describe how toric geometry gives an efficient
and systematic  framework for generating a basis for the 
relative cohomology group $H^3(X,\Z)$ and a system of 
differential equations associated to it. 

\subsec{Relative cohomology in toric \CY $X$}
We consider a toric \CY $d$-fold 
defined by a 2d LSM with $N$ matter fields $y_n$, whose mirror $X$
can be described by a 
superpotential of the form\foot{We use a tilde
to distinguish the 2d superpotential $\tW$ from the 
superpotential in 4d space-time. For 
background material on the following definitions, see
\doubref\HV\HIV.} 
\eqn\deftoric{
\tW=\sum_n a_ny_n,\qquad \prod_n (y_n)^{\ll a_n}=1, 
\qquad a=1,\dots, h^{d-1,1}(X)\ .
}
Here the $\ll a$ are
$N-d=h^{d-1,1}(X)$ integral vectors that 
define relations between the fields $y_n$. Moreover the coefficients  
$a_n$ are $N$ complex parameters that specify (with some
redundancy) the complex structure of $X$.

In view of the example that we will present further below,
we will here discuss non-compact 
Calabi-Yau's $X$, for which the $y_n$ are variables
in $\CC^*$ and for which we do {\it not} impose $\tW=0$ 
\HV.  
We denote the true $\CC$ variables by $u_n=\ln(y_n)$. The
following discussion can be also adapted to 
compact \CY manifolds, where one imposes $\tW=0$ and where the 
$y_n$ are variables with values in $\CC$.  

The relations in \deftoric\ can be solved in terms of 
$d$ of the $N$ variables $y_n$. In the following we 
order the variables $y_n$ such that these $d$ variables are 
$y_i$, $i=0,..,d-1$, whereas variables $y_{\hx k}$ 
with hatted indices, $\hx k \geq d$, denote the variables 
eliminated in this step.
The resulting superpotential in the $d$ variables $y_i$ can be written as
$$
\tW(y_i)=\sum_{n=0}^{N-1} a_n \bigg(\prod_{i=0}^{d-1} y_i^{v^i_n}\bigg)\ ,
$$
where $v^i_n$ are $d$ null vectors of the matrix $l$
defined by the choice of coordinates $y_i$. They 
satisfy  $\sum_{n} \ll a _n v_n^i=0$ for all $a$ and $i$.

In the patch parametrized by the variables $u_i=\ln(y_i)$,
the holomorphic $d$-form can be written as \doubref\Cec\HV
\eqn\omtor{
\Om=\Big(\prod_{i=1}^d du_i \Big)\, e^{-\tW}.
}
The set of $N$ derivatives:
\eqn\logder{
\fc{\p}{\p  a_n}\, \Om = -y_{n}\, \Om,\qquad n=0,...,N-1,
}
can be reduced to a basis of $h^{d-1,1}$ preferred derivatives 
in the following way. A natural choice for this basis is to keep 
the derivatives $\p/\p a_{\hx k}$  
with respect to the parameters  $a_{\hx k},\, \hx k\geq d$
(which are the coefficients of the variables 
that have been eliminated in the first step). 
The derivatives 
$$
\tW_i =  \fc{d}{du_i}\ \tW = \sum_{n=0}^{N-1} a_n v_n^i\, y_n,
$$
provide $d$ equations to eliminate the $d$ variables 
$y_i$ for $i< d$ in the expression \logder\ for $\p/\p a_i$ in
favor of the derivatives $\tW_i$. An expression involving
$\tW_i$ can then be further simplified by noting that the form
$$
\Big(\prod du_j\Big)\, \tW_i e^{-\tW} = d(\om_i),\qquad \om_i = (-)^i
\Big(\prod_{j\neq i} du_j\Big)\, e^{-\tW}
$$
is exact in the absolute cohomology on $X$.\foot{Although
our presentation illustrates the underlying idea of the computation, 
this interpretation is really oversimplified. As we 
are using here the parameters $a_n$, some of which are redundant 
by $\CC^*$ rescalings of the coordinates, the equations for $\Om$
written in this section 
should be interpreted in the $\CC^*$ equivariant 
cohomology on~$X$.}
The $d$ derivatives $\fc{\p}{\p a_{i}}\,\Om$ for $i < d$ can thus be
expressed in terms of the $h^{d-1,1}$ independent derivatives
$\Om_{n} = \fc{\p}{\p a_{n}}\Om$ for $n \geq d$ as follows:
\eqn\imp{
a_i\fc{\p}{\p a_i}\Om = d(\om_i) -\sum_{n\neq i } v_{n}^i a_{n}
\Om_{n}\,.
}
Note that \imp\ can be used to express any 
product of derivatives in the $a_n$
in terms of multiple derivatives with respect to the $h^{d-1,1}$ chosen 
derivatives $\fc{\p}{\p {a_\hx k}}$,
plus exact pieces that are derivatives
of the exact piece in \imp. Thus the problem of 
rewriting an arbitrary form obtained by acting with 
derivatives on $\Om$, in terms of elements of a chosen basis for $H^d(X)$ 
plus exact pieces, is already completely solved at this step. 

This should
be contrasted with Griffiths' reduction method \doubref\Grif\LSW,
which applies to the general, non-toric case but 
is technically by far more complicated.
The key point is that the toric parameters
$a_n$ linearize the moduli problem, a fact that leads to a
well-known simplification in the derivation of the Picard-Fuchs
equations for the absolute cohomology $H^d(X)$. 
The equations \imp\ describe an similar simplification for
the treatment of exact pieces, which are relevant for the 
derivation of the differential equations for the relative 
cohomology group $H^d(X,\Z)$.
Specifically these exact pieces become important in open string
backgrounds, where the extended chiral ring $\Roc$
lives in the relative cohomology $H^d(X,\Z)$ and where
one has to keep track of boundary terms. 

{}From now on we specialize to the case $d=3$, i.e., a \CY 3-fold,
and to a single $D$-brane on top of it.
In order to define the relative cohomology group $H^3(X,\Z)$, 
we need to specify an appropriate family of 
hypersurfaces $\Z(\zh)$ in the toric \CY $X$ that slice 
the relevant special Lagrangian 3-chains. 
We will use the following ansatz, whose form will be derived
in sect.~5.3,
\eqn\moddefii{\Z(\zh):\ y_0= \zh \, y_1 \big(\fc{y_1}{y_2}\big)^\nu,\qquad
\hx z= \fc{a_0}{a_1}\big(\fc{a_2}{a_1}\big)^{\nu},}
where $y_i,\, i=0,1,2$, denote the coordinates on the 3-fold $X$.
Moreover $\hx z$ is the open string modulus that specifies the position 
of the hypersurface $\Z(\hx z)$ that passes through the boundary 2-cycle
wrapped by the D-brane.
As argued later, with the above ansatz
the parameter $\zh$ is one of the good algebraic coordinates
on the the moduli space $\MM$. 
That is, $\zh\to 0$ defines a classical limit in which
the associated flat coordinate has the leading 
behavior $\hat t\sim\fc{1}{2\pi i }\ln(\zh)$.

\subsec{A system of GKZ equations for the relative cohomology}
{}From the above considerations, one may obtain a complete system 
of differential equations for the relative cohomology as follows.
The holomorphic 3-form $\Om$ on the \CY manifold $X$ 
satisfies a system of differential equations, a so-called 
GKZ \gkz\ system 
\eqn\egkz{
\cx L_a \ \Om = \Big(~\prod_{\ll a_i>0}\p_{a_i}^{\ll a_i}-
\prod_{\ll a_i<0}\p_{a_i}^{-\ll a_i}~\Big)\ \Om =0\,,
\qquad a=1,\dots, h^{2,1}(X)\ .
}
These equations follow straightforwardly from the explicit expression 
\omtor\ for $\Om$. 
An important point is that the differential equations \egkz\ 
continue to hold in the
relative cohomology $H^3(X,\Z)$, if the canonical vacuum is defined  by
$\vx \Om=(\Om,0)$ as in \defcom.
However the content of these
equations is different; first they now describe differential
equations that involve both, the open and closed string moduli
$z_A=(z_a,\hx z)$, defined respectively by \moddefii\ and
\eqn\defza{
z_a\ =\ \prod_n (a_n)^{\ll a_n}\,,\qquad a=1,\dots,h^{2,1}(X)\ .
}
The $z_a$  are good local 
parameters for the complex structure, and are
invariant under those $\CC^*$ rescalings of the $y_i$ 
that preserve the form of the LG potential for $X$.

Second, derivatives of $(\Om,0)$ 
produce terms that, despite being exact in the absolute 
cohomology $H^3(X)$, generate non-trivial elements in the 
relative cohomology $H^3(X,\Z)$ on the boundary $\Z$, as discussed in sect.~5.1. 
Thus, the differential operators $\cx L_a$ now represent linear constraints 
on the space spanned by the linearly independent
elements in $H^3(X,\Z)$ (which represents the open-closed chiral ring
$\Roc$).

There are additional differential equations needed to augment
the GKZ system \egkz\ to a complete system of differential
equations for the periods of the relative cohomology $H^3(X,\Z)$.
The origin of these equations is obvious: They 
are the differential equations satisfied by the holomorphic 2-form 
\eqn\omdef{\om = du_1du_2\,e^{-\Wt_{\Z}},\hskip 12pt
\Wt_{\Z}(y_1,y_2) = \Wt|_{\Z}\ ,}
on the hypersurface $\Z$ defined by the LG potential $\Wt_\Z$.
It is straightforward to see from these expressions that 
$\om$ satisfies the following differential equations:
\eqn\egkzii{\cx L_a\, \om = 0,\hskip12pt \widehat{\cx L}\, \om = 0,}
where $\widehat{\cx L}$ is a differential operator of the 
form \egkz\ defined by the vector% 
\eqn\lhatdef{
\hx l = (1,-\nu-1,\nu,0,...,0).
}
Moreover, eqs.\moddefii\ and \defza\ imply that
$$
\hx \theta\, \Om = \zh\fc{d}{d\zh}\, \Om = \rmx{const.} \ \om\ .
$$
It follows that the following differential operators provide a
complete system that determines the periods of the relative cohomology:
\eqn\egkzf{
\eqalign{
\cx L_a\, \Om &= 0\,,\qquad a=1,\dots,h^{2,1}(X)\,,\cr
\widehat{\cx L}\, \hx\theta\, \Om &= 0\,.
}}

In passing we note that the above discussion extends the open-closed
string dualities found in \PM,
from the special case $\nu=0$ to all values of $\nu$. 
Specifically, there exists a ``dual'' 
closed string compactification on a \CY 4-fold $M$ without branes 
for any of the open-closed string geometries considered above. 
In particular, the holomorphic
potentials $\cx W_K$ of the open-closed string are reproduced
by the genus zero partition function $\cx F_0$ of the 
closed string on $M$. This implies
a quite amazing coincidence of the ``numbers'' of sphere and disc
instantons counted by the A-model for the D-brane geometry on the mirror of
$X$, and the ``numbers'' of sphere instantons computed by the 
A-model on the mirror of the 4-fold $M$. As the necessary
modifications of the 4-fold geometries for $\nu\neq 0$ are straightforward when starting
from the 4-folds described in \PM, we will not go into the
details, except for a brief comment in sect.~6.

\subsec{Degree $\nu$ hypersurfaces $\Z$, framing ambiguity, 
winding sectors  and open-closed string  moduli}
It remains to justify the ansatz \moddefii\ for the 
family of hypersurfaces $\Z$  in the toric \CY $X$ that slice 
the relevant special Lagrangian 3-chains. 

In fact, the definition of the family of hypersurfaces $\Z\subset X$
is parallel to the definition of the \CY $X$
itself. To start with, note that the relations defined by the $h^{2,1}$
charge vectors $\ll a$ define homogeneous ``hypersurfaces''\foot{This terminology is not quite precise, as will be discussed momentarily.} $H_a$
and $X$ is defined on the intersection $\cap_a H_a$ 
of these $h^{2,1}$ hypersurfaces. 
One may associate moduli to these hypersurfaces as in \defza, by writing
\eqn\moddefi{
H_a:\ \prod_n (y_n)^{\ll a_n}=z_a, \hskip 30pt
z_a = \prod_n (a_n)^{\ll a_n}.
}
A modulus $z_a$ is, by definition, invariant under the
$\CC^*$ action that preserves the hypersurface $H_a$, but on
the other hand parametrizes a $\CC^*$ action that moves the hypersurface
$H_a$ in the ambient space. Specifically, the (closed string) 
moduli $z_a,\, a=1,...,h^{1,2}$,
provide coordinates on the moduli space of complex structures
on $X$ that are invariant under those $\CC^*$ rescalings of
the coordinates~$y_i$ that preserve the LG potential. 

The ansatz \moddefii\ for the hypersurface $\Z$
$$
\Z(\hx z):\ y_0=\hx z \, y_1 \big(\fc{y_1}{y_2}\big)^\nu,\qquad
\hx z= \fc{a_0}{a_1}\big(\fc{a_2}{a_1}\big)^{\nu},
$$
is of the same general form as \moddefi.
It is subject to the following two constraints: $i)$ it is
homogeneous in the coordinates $y_i$; 
$ii)$ in the limit $\hx z\to 0$,
$\Z(\zh)$ approaches the hypersurface $y_0=0$ with 
an asymptotic behavior linear
in $\hx z$.\foot{Of course the choice of singling out $y_0$
is a convention, taken for simplicity of notation. There are
similar phases of D-branes located in other patches of $X$,
described by hypersurfaces $\Z$ whose classical limit is $y_n=0$ 
for some other $n$.}

Let us now explain the origin of these conditions. Although it seems natural,
the homogeneity of the ansatz for $\Z(\zh)$ is
not really obvious. In fact this constraint is mirror
to that derived in \AVi\ on 
a {\it supersymmetric} D-brane in the A-model. 
This constraint imposes the vanishing of the sum over the components
of $\hx l$ in \lhatdef.
Note that in the B-model that we are
considering, it implies the homogeneity of the equation and 
is indeed nothing but the condition that
the intersection $\Z\cap X$ is a (possibly singular)
\CY manifold of two dimensions\foot{For
$\nu=0$, the intersection $\Z\cap X$ is easily recognized as the
standard form for the mirror of an ALE space, as 
illustrated further in sect.~6.2. }.

The second constraint, item $ii)$, needs a little more explanation.
It is related to a fundamentally new geometric structure in the 
open-closed string compactification, which is related to the logarithmic multi-valuedness 
of the good coordinates $u_i = \ln(y_i)$.%
\foot{The following discussion is very similar 
to the argument for the identification of the open string 
modulus with a D-brane tension in ref. \AVii. It
is more concrete in the sense that it identifies the 
``framing ambiguity'' discussed there, with 
the degree of the hypersurface $\Z$.}
First note that an equation like \moddefi\ does not really 
define a hypersurface, but a lattice of hypersurfaces 
in the true coordinates $u_i$, one for each of 
the sheets of the logarithm. Specifying a Dirichlet boundary 
condition for an open string in $X$ includes therefore the selection of 
a particular sheet of the logarithm $u_i=\ln(y_i)$. This implies that there
will also be extra winding states in the open string sector that interpolate
between the hypersurfaces $\Z$ on different log sheets \AVii. In contrast,
the observables in the closed string 
B-model are related to absolute homology cycles in $X$ and thus
live on a single, connected sheet.

The lattice generated by the log sheets for 
the open string sector should be seen as the way the 
\CY manages to meet the expectations from T-duality. As is well-known,
the Wilson line on a D-brane wrapped on $S^1$ is mapped by T-duality to 
a position on the dual $S^1$. The phase of the 
open string modulus for the A-type geometry which is mirror to our B-model,
is precisely a Wilson line on the type IIA D-brane \doubref\WitCS\vafaext. This 
phase needs to be mapped to a position on a circle in the \CY
$X$ for the B-model, but there are no $S^1$'s in the \CY manifold.\foot{Note 
that the crux of the T-duality is that the gauge degree of freedom 
on the D-brane becomes geometric in the B-model.} This $S^1$
structure, which is supposedly seen in the open string sector 
but not in the closed string sector, is realized by the lattice of 
log sheets which appear in the very definition of the manifold $X$. 
T-duality suggests that a similar logarithmic structure could be also 
relevant for compact \CY manifolds with open string sectors. 

The constraint $y_0\sim \zh$ near $\zh=0$ in \moddefii\ 
is explained
by the above considerations as follows. Near a classical limit,
where the instanton corrections to the superpotential are suppressed,
the good local open string modulus $\zh=0$ will satisfy
$\hx z \sim e^{2\pi i \hx t}$. In this limit,
the classical mirror A-model geometry is valid and we can identify the 
real part of $\hx t$ with the Wilson line on the A-type brane. An 
integral shift $\hx t\to \hx t+2\pi$ of the Wilson line is mirror 
to a full rotation on the T-dual circle, and it must map the hypersurface 
$\Z$ to ``the same'' hypersurface, up to a change of the sheet of the
logarithm induced by $y_0 \to e^{2\pi i } y_0$.

As this is a classical argument which is valid in the limit $\hx z \to 0$,
it fixes the asymptotic behavior $y_i \sim \hx z$, for a 
hypersurface $\Z$ that is classically defined by $y_i=0$; it 
does not fix the degree of the hypersurface $\Z$ in \moddefii,
which is given by the integer $\nu$.

The integer $\nu$ in the definition \moddefii\ of $\Z$ 
is in fact an geometric invariant of our B-model
which represents the ``framing ambiguity'' discovered in \AVii.
It is related to the choice of a framing for a knot in the 
the Chern-Simons theory on the A-type mirror brane.

The geometric identification of the framing number $\nu$
in the D-brane geometry of the A-model has been given in \yaip, 
as the number of times the boundary of an world-sheet instanton
ending on the A-type brane wraps around the D-brane in transverse space.
This wrapping number can be specified by choosing a $U(1)$ action
on the three coordinates of the A-model, relative to the $U(1)$ defined
by the boundary of the world-sheet instanton. 

{}From the equation
\moddefii\ we see that the interpretation of the framing number as
the degree of the hypersurface $\Z$ is precisely the mirror of
this statement. Rewriting the constraint as $\hx z= (y_0/y_1)(y_2/y_1)^{\nu}$,
one observes that the phase of $\hx z$ (identified with the $U(1)$ acting
on the $S^1$ boundary of an world-sheet instanton) 
defines a particular $U(1)$ action $T$ on
the phases of the coordinates $y_i$. The integer $\nu$ 
specifies the orientation of the $U(1)$ action $T$ w.r.t to the
classical definition of the $U(1)$ on the D-brane, which is 
the phase of $y_0/y_1$.

%%%%%%%%%%%%%%%%%%%%%%%%%%%%%%%%%%%%%%%%%%%%%%%%%%%%
\newsec{A case study: D5-branes on the mirror of the 
non-compact \CY $X=\cx O(-3)_{\IP^2}$}
%%%%%%%%%%%%%%%%%%%%%%%%%%%%%%%%%%%%%%%%%%%%%%%%%%%%

\noi
To illustrate the somewhat abstract ideas 
presented above, we will now study a particular D-brane geometry 
in some detail. Specifically, we consider the non-compact
\CY $X$ that is mirror to the canonical bundle on $\IP^2$.
The superpotential for a D5 brane in this geometry has 
been computed already in \refs{\AVii,\PM,\LM}.

The single integral vector for the LSM describing this 
non-compact Calabi-Yau manifold is given by
$l^{(1)}=(-3,1,1,1)$. Accordingly, the mirror geometry
is captured by the following Landau-Ginzburg (LG) superpotential:
\eqn\PtwoWdef{
\Wt=a_0y_0+a_1y_1+a_2y_2+a_3y_3,\qquad {\rm where\ \ }
y_3=y_0^3/y_1y_2.
}
The $D$-brane geometry is specified by the 
``hypersurface'' $\Z$, defined as in \moddefii.
The classical limit $y_0=0$ for $\Z$ 
corresponds to a $D$-brane in the ``outer phase''
in the nomenclature of ref.\ \LM. In order to describe geometries with
general $\nu$, it is convenient to 
change variables $y_0\to y'_0\, (y_1/y_2)^{\nu}$.
The surface $\Z$ can then be represented by setting $y'_0=y_1$
in the superpotential

$$
\Wt(\nu) = a_0y'_0y_1^\nu y_2^{-\nu}
+a_1y_1+a_2y_2+a_3y^{'\, 3}_0y_1^{3\nu-1}y_2^{-3\nu-1}.
$$

\noi
The derivatives of the holomorphic (3,0) form $\Om$ defined
as in \omtor\ obey the relations

\eqn\imp{\eqalign{
a_0\p_{a_0}\Om &= d\om_{0'} - 3 a_3\p_{a_3}\Om
\cr
a_1\p_{a_1}\Om &= d\om_1 -\nu\, d\om_{0'} + a_3\p_{a_3}\Om
\cr
a_2\p_{a_2}\Om &= d\om_2 +\nu\, d\om_{0'} + a_3\p_{a_3}\Om
\ ,
}}

\vskip4pt
\noi where $\om_{0'}=d(du_1du_2e^{-\tW(\nu)})$, and similarly 
for the other $\om_i$. The forms $d\om_i$ are exact in $H^3(X)$.
However, as emphasized before, they contribute in 
chain integrals with non-zero boundary components, and
represent non-trivial elements in the relative cohomology.
Specifically, under ``partial integration'' one has 
$\om_2\vert_\Z= 0$, $\om_0\vert_\Z=-\om_1\vert_\Z=\om$, 
where $\om$ is the holomorphic $(2,0)$-form on $\Z$:
\eqn\Zomdef{
\om = du_1du_2e^{-\WZ(\nu)}.}
By iteratively applying \imp\ and reducing the exact pieces
to forms on $\Z$, one may express any derivative
of $\Om$ in terms of $a_3$-derivatives of $\Om$ and $\om$. 

In eq.~\Zomdef, the expression $\WZ(\nu)$ denotes the restriction 
of the superpotential $\Wt$ to $\Z$:
\eqn\WZdef{
\WZ(\nu)\ =
a_0 y_1^{\nu+1} y_2^{-\nu}+a_1y_1+a_2y_2+a_3y_1^{3\nu+2} y_2^{-3\nu-1},
}
Note that for a hypersurface $\Z$ of degree 1, that is, for the hyperplanes
defined with $\nu=0,-1$,
this turns into the standard LG superpotential for the mirror of 
an ALE space with $A_1$ singularity, however with an unusual
parametrization of the moduli. E.g., for $\nu=0$ one finds
\eqn\WALEdef{
\WZ(0)=(a_0+a_1)y_1+a_2 y_2 + a_3 y_3,\qquad {\rm where\ \ }
y_3=y_1^2/y_2.
}

Note also that after a rescaling $y_i\to a_i^{-1}y_i$, the potential $\Wt(\nu)$
and the constraint for $\Z$ depend only on the algebraic coordinates
$z_A=(z_1;z_2)$, where
\eqn\exvardef{
z_1 = \fc{a_1a_2a_3}{a_0^3},\qquad \z_2=\hx z=\fc{a_0}{a_1}
\big(\fc{a_2}{a_1}\big)^\nu,
}
represent the bulk modulus and the brane modulus, respectively.

In the following two sections, we will discuss separately the cases 
where $\Z$ is either a hyperplane, or a hypersurface of higher degree.

\subsec{The hyperplane case, $\nu=0$}

We should mention here that the motivation for this sub-section is {\it not}
the derivation of the differential equations for the flat coordinates
and the superpotential; these equations have been written down in
full generality in sect.~5.2. Rather, the following computations demonstrate
how the various 3-forms on the \CY $X$, and the 2-forms on the ALE space
defined by the ``hyperplane'' $\Z$, nicely fit together to a basis for
the relative cohomology group $H^3(X,\Z)$. This will allow to give an
explicit representation of the open-closed chiral ring, and to write down its
structure constants. From these we will then determine, via the
matrix differential equations \pfoc, the relative period matrix.

The starting point is to express arbitrary derivatives w.r.t.~the moduli
of the holomorphic $(3,0)$ form, in terms of a suitable basis
of differential forms spanning $H^3(X,\Z)$. 
By iteratively applying \imp\ (with $\nu=0$), we can express any derivative
of $\Om$ in terms of $a_3$-derivatives of $\Om$ and $\om$. 
We have collected some sample expressions in the following table:

\vbox{
\vskip 10pt
\eqn\modids{
\vbox{\offinterlineskip\tabskip=0pt\halign{
\strut
~$#$~\vrule
&~$#$~&~$#$~&~$#$~\vrule
&~$#$~&~$#$~&~$#$~\vrule
&~$#$~&~$#$~\vrule
&~$#$~\vrule\cr
&{a_3}^3\Om'''&{a_3}^2\Om''&a_3\Om'
&{a_0}^2h^2\om''&a_0a_1h^2\om''&{a_1}^2h^2\om''
&a_0h\om'&a_1h\om'
&\om\cr
\noalign{\hrule}
\theta_1& &&1&&&&&&\cr
\theta_{\two}& &&&&&&&&1\cr
\noalign{\hrule}
\theta_1^2&&1&1&&&&&&\cr
\theta_1\theta_{\two}&&&&&&&1&1&\cr
\theta_{\two}^2& &&& &&&1&3&\cr
\noalign{\hrule}
\theta_1^3&1&3&1&&&&&& \cr
\theta_1^2\theta_{\two}&&&&1&2&1&1&1&\cr
\theta_1\theta_{\two}^2&&&&1&4&3&1&3&\cr
\theta_{\two}^3&&&&1&6&9&{a_0\over a_0+a_1}&{9a_1+4a_0\over a_0+a_1}&\cr         
\noalign{\hrule}
}}}}
\noi 
Here a prime denotes $\p\over\p {a_3}$, $h= {a_3\over a_0+a_1}$, 
and $\theta_A=z_A\p_{z_A}$ are the logarithmic derivatives 
with respect to the good local coordinates \exvardef\ (for $\nu=0$)
of the combined open-closed string moduli space.

The list of the forms displayed on the top of the table can be
further reduced, by making use of relations between the derivatives 
$\p/\p a_n$ that
originate from the specific form of $\Wt$ and $\WZ$ for $\nu=0$.
The first such relation is the GKZ system \egkz\ for 
$X=\cx O(-3)_{\IP^2}$:
$$
{\cal L}_1\Om\ =\ 
\big(\p_{a_1}\p_{a_2}\p_{a_3}-{\p^3_{a_0}}\big)\Om\ =\ 0\ .
$$
It allows one to eliminate  $\Om'''$ in favor of lower-order derivatives.

As indicated earlier, there are additional
differential equations \egkzii\ satisfied by the $(2,0)$ form 
$\om$ \Zomdef\ on the $A_1$ ALE space, $\Z$. 
The differential operator associated with $\hx l$ is 
\eqn\gkzi{
\widehat{\cal L}_1\,\om\ =\ (\p_{a_1}-\p_{a_0})\,\om\ =\ 0\,,
}
It expresses the redundancy of the parametrization of the coefficients
in $\WZ$. Using the trivial identities
$$
\p_{a_0}= \fc{1}{a_0}(\theta_{\two}-3\theta_1),\qquad
\p_{a_1}= \fc{1}{a_1}(\theta_1-\theta_{\two}), 
$$
it gives rise to the second PF operator in
eq.(6.11) below. 

Next we consider a certain linear combination of the differential
relations in \egkzii, corresponding to the following vector of relations:
$l^{(1)}-\hx l$; this gives an alternative viewpoint of the
geometric meaning of the differential operators.  
In fact, this second differential operator annihilating $\om$ describes the
ordinary GKZ equation for the ALE space with $A_1$
singularity. The latter is defined by the 
charge vector $l=(-2,1,1)$, leading to the relation
\eqn\gkzii{
\widehat{\cal L}_2\,\om\ =\ (\p_{\ti a_2}\p_{\ti a_3}-{\p_{\ti a_1}}^{\!\!2})\,\om\ =\ 0\ .
}
{}From \WALEdef, 
the parameters $\tx a_n$ for the $A_1$ singularity are related to the 
complex structure parameters for $X$ by $\ti a_1=a_1+a_0$,  
$\ti a_2=a_2$,  $\ti a_3=a_3$. 
The differential equation \gkzii\ can be used to 
eliminate $\om''$ in the table in favor of $\om'$:
$$
a_3\om'' = \Big({6 \ti z-1\over 1-4 \ti z}\Big)\om' \,,
$$ 
where $\ti z=\ti a_2\ti a_3/{\ti a_1}^2$.

The three differential equations thus
allow one to reduce the forms in the table to
a minimal basis given by $\{\Om, \Om',\Om'',\om,\om'\}$.

{}From now on it will be more convenient to
switch to the good local variables $z_A$ defined in \exvardef\
(with $\nu=0$). In terms of
these variables we obtain, after making some convenient linear 
combinations, the following system of differential equations:
\eqn\PFdef{
\eqalign{
{\cal L}_1\Om\ &=\ 0,\cr
\widehat{\cal L}_{1,2}\,\om\ &=\ \widehat{\cal L}_{1,2}
\,\theta_{\two}\,\Om\ =\ 0\ ,
}}
with
\eqn\Lops{
\eqalign{
{\cal L}_1&\ =\ 
{{\xt(1)}^2}\left( \xt(1) - \xt(\two) \right)  +
  \zo\left( 3\xt(1) - \xt(\two) \right)
   \left( 1 + 3\xt(1) - \xt(\two) \right)
   \left( 2 + 3\xt(1) - \xt(\two) \right)\ ,
\cr
\widehat{\cal L}_1&\ =\ 
\left( 3\xt(1) - \xt(\two) \right) -
\zt\left( \xt(1) - \xt(\two) \right)\ ,
\cr
\widehat{\cal L}_2&\ =\ 
\left(  \zt-3 \right) 
((1 - \zt)^2+ 4{{\zt}^3}\zo)
 {{\xt(\two)}^2}
+2\zt\left( 1 - \zt - 9{{\zt}^2}\zo + {{\zt}^3}\zo \right) 
   {{\xt(\two)}}\,.
}}
As discussed in sect.~5.2, one of the operators, i.e.,
$\widehat{\cx L}_2$, is in fact redundant.
A complete system of differential operators for $\Om$, 
namely $\{\cx L_1,\widehat{\cx L}_1
\theta_2\}$,  had already been obtained in \doubref\PM\LM\foot{See also
\govi\ for a closely related discussion.}.

Note that due to the simple form of the second and the third operators, 
one can easily write down the general solution for the system 
$\widehat{\cal L}_{1,2}\,\om=0$, which yields
the periods of the $(2,0)$ form $\omega$. It 
is given by the logarithmic derivative of the space-time superpotential:
\eqn\ThW{
\theta_{\two}\cW\equiv{\rm const.}\,\log\Big[  {1-z_{\two}+\sqrt{(z_{\two}-1)^2+4 z_1 {z_{\two}}^3}\over{\sqrt{4
z_1 {z_{\two}}^3}}} \Big]+{\rm const.}
}

The operators \Lops\ enable us to 
express any derivative of $\Om$ in terms of 
the following minimal basis for the relative cohomology $H^3(X,\Z)$:
\eqn\formbasis{
\eqalign{
\vec\pi(z)\ &=\ \big\{(\Om,0),\,(\xt(1)\Om,0),\,(0,\om),\,({\xt(1)}^2\Om,0),\,
(0,\xt(1)\om)\big\}^t\ .
}}
The action of single logarithmic derivatives $\theta_A$ on $\vec\pi$ 
can be expressed in terms of the linear system \betterbasis,
i.e.,
\eqn\GMsystem{
D_A\cdot \vec\pi(z)\ = \ 0\ , \qquad A=1,2\ ,
}
where 
\eqn\Cmatr{
\eqalign{
D_1&={\bf 1}\xt(1)-
\pmatrix{ 0 & 1 & 0 & 0 & 0 \cr 0 & 0 & 0 & 1 & 0 
\cr 0 & 0 & 0 & 0 & 1 \cr 0 & {
     -\frac{6\zo}{\Delta_1}} & {\frac{2\zo}{\Delta_1}} & {-\frac{27
      \zo}{\Delta_1}} & {\frac{\zo B}
{\Delta_1  \Delta_2 \Delta_3}} \cr 0 & 0 & 
   0 & 0 & {-\frac{2{{\zt}^3}\zo}
    {\Delta_2}}  }\cr
D_{\two}&={\bf 1}\xt(\two)-
\pmatrix{ 0 & 0 & 1 & 0 & 0 \cr
 0 & 0 & 0 & 0 & 1 \cr 0 & 0 & 0 & 0 & {\frac{\zt-3}
    {\Delta_3}} \cr 0 & 0 & 0 & 0 & {-\frac{2{{\zt}^3}\zo}
    {\Delta_2}} \cr 0 & 0 & 0 & 0 & {\frac{2
      \left(3- \zt\right) {{\zt}^3}\zo}{\Delta_3
      {\Delta_2} }} \cr  }\ .
}}
Here
$
B= -9 + 31\zt - 37{{\zt}^2} + 
{{\zt}^3}( 17 - 18\zo)+2
{{\zt}^4}(11\zo-1)\ ,
$
and the $\Delta_i$ are discriminant factors associated with the
PF system:\foot{Note that $\Delta_2=0$ is reminiscent
of the splitting of classical singularities in $N=2$ SYM theory.
It would be interesting to investigate this from the view point of
non-perturbative brane physics; for example, 
in terms of domain walls becoming tensionless.}
\eqn\Deltadef{
\eqalign{
\Delta_1\ &=\ 1 + 27\zo,\cr 
\Delta_2\ &=\ (1 - \zt)^2+ 4{{\zt}^3}\zo,\cr
\Delta_3\ &=\ \zt-1.
}}

\noi The matrices \Cmatr\ give an explicit representation of
the maps in the following diagram of the variations
of mixed Hodge structures:
%ddd
\def\Thetap{\theta^{+1}}
\eqn\vmhsEx{
\xymatrix{
(\Om,0)\ar[r]^{\ss \Thetap_1}\ar[dr]^{\ss \Thetap_2}&
(\theta_1\Om,0)\ar[r]^{\ss \Thetap_1}\ar[dr]^{\ss \Thetap_2}&
(\theta_1^2\Om,0)\ar[dr]^{\ss \Thetap_1\!\!\!,\ \Thetap_2}\\
&(0,\om)\ar[r]^{\ss \Thetap_1\!\!\!,\ \Thetap_2}&
 (0,\theta_1\om)\ar[r]^{\ss \Thetap_1\!\!\!,\ \Thetap_2}& 0}}
Here the superscript on $\theta_a^{+1}$ denotes the grade one piece in the 
derivative. 
Note that the diagram truncates at grade two 
because of the non-compactness of both
of the 3-fold $X=\cx O(-3)_{\IP^2}$ and the embedded ALE space $\Z$.

A consistency check on the matrices \Cmatr\ is that they satisfy
\eqn\flatnessEx{
\big[D_A\,,\,D_B\big]\ =\ 0\ ,
}
and this explicitly demonstrates the flatness and 
integrability of the combined closed and open string moduli space. 
As discussed in sect.~4.2, we know that in general:
$D_A=\nabla_A-C_A=\theta_A-\GMC_A-C_A$. Moreover, we know that the flatness
property \flatnessEx\ allows us to introduce flat coordinates, $t_A$, and
go to a gauge in  which the Gauss-Manin connection $\GMC_A$ 
vanishes 
(the ring structure constants $C_A$ are distinguished by the property
of being strictly upper triangular, 
{\it i.e.}, having zeroes on the diagonal).   

This is easy to explicitly verify in the present example, 
by making use of the solutions of
the differential system \Lops. As discussed before, 
a classical limit is defined by $z_A=0$ and the leading 
behavior of the flat coordinates is $t_A=\ln(z_A)+\cx O(z_A)$. 
A power series in the $z_A$ then represents an 
instanton expansion in the exponentials $q_A=\exp(2\pi i t_A)$,
which is invariant under the shifts of the real parts, $t_A \to t_A + 2\pi$.
The exact solutions to the differential operators \Lops\ with 
the requisite leading behavior are known and given by 
\eqn\mirrormap{
t_1(z_\dd)\ =\  \log(z_1)-3 S(z_1)\ ,\qquad 
t_2(z_\dd)\ =\ \log(z_{\two})+S(z_1)\ ,
}%
where
\eqn\Bdef{
S(z_1)=-\sum_{{n_1>0}}
{(-)^{n_1}(3n_1-1)!\over {n_1!}^3}
 {z_{1}}^{n_1}
\ .}
One may check that after transforming to the coordinates $t_A$, 
the lower-triangular parts of $D_A$ indeed vanish. There are still some 
degree zero terms in the connection due to the fact that we have 
not properly fixed the normalization of the forms in \formbasis.  
Moreover there is the further
ambiguity in that forms of the same degree can mix,
so the complete flattening of the 
connection,  including all of its block-diagonal parts, 
will require rescalings of the basis of differential forms  
along with a careful choice of linear combinations.
Ultimately we are led to the following flat basis for the relative
cohomology:
\eqn\flatformbasis{
\vec\pi(t) =\big\{(\Om,0),\,(\del_{\xo}\Om,0),\, (0,\eta),
(f\p^2_{t_1}\Om,-f\, h{\del_{\xo}}\eta),\, 
(0,g\,\del_{\xo}\eta)\big\}^t\
\in\ H^3(X,\Z)\ ,
}
where we have defined $d\eta \equiv \del_{\xs}\Om$, and the
rescalings are given by $f=1/\del_{\xo}^3\cF$,
$g=1/\del_{\xs}\del_{\xo}\cW$, and $h=g \del_{\xo}^2\cW$;
the functions $\cF$ and $\cW$ will be defined below.  
This period vector is a solution of the completely flattened 
Gauss-Manin system: 
\eqn\flatGM{
\Big({\partial\over \partial t_A}\ -
\ C_A(t)\,\Big)\cdot \vec\pi(t)\ = 0\ ,
\qquad A=1,2. 
}
where the open-closed chiral ring structure constants have
been obtained from \Cmatr\ and the 
mirror map \mirrormap\ as follows:
\eqn\ringele{
\eqalign{
C_1(t)&=\pmatrix{ 0 & 1 & 0 & 0 & 0 \cr
0 & 0 & 0 & \del_{\xo}^3\cF & \del_{\xo}^2\cW \cr
0 & 0 & 0 & 0 & \del_{\xs}\del_{\xo}\cW \cr
0 & 0 & 0 & 0 & 0\cr
0 & 0 & 0 & 0 & 0 \cr  }\ ,
\cr
C_{\two}(t)&=\pmatrix{ 
0 & 0 & 1 & 0 & 0 \cr 
0 & 0 & 0 & 0 & \del_{\xs}\del_{\xo}\cW\cr 
0 & 0 & 0 & 0 &  \del_{\xs}^2\cW \cr
0 & 0 & 0 & 0 &  0 \cr 
0 & 0 & 0 & 0 & 0 \cr  }\ .
}}
As expected, $\cF=\cF(\xo)$ as computed in this way turns out
to precisely coincide with the known bulk $\cx N=2$ prepotential 
associated with $X=\cx O(-3)_{\IP^2}$, which is of the form\foot
{The $k^{th}$ polylogarithm is defined by 
$Li_k(q)= \sum_{n>0} q^n/n^k$ for $k \geq 1$.}
$\cF(t_1)=-{1\over18}{t_1}^3+
\sum_{n_1}N^{(1)}_{n_1}Li_3(e^{2\pi i n_1t_1})$;
the precise values of the sphere instanton coefficients $N^{(1)}_{n_1}$ 
are known but are not important here.

Moreover, $\cW=\cW(t_A)$ coincides with the known superpotential on the
world-volume of the D-brane, for which the instanton expansion 
is of the form: 
\eqn\superpot{
\eqalign{
\cx W(t_A)\ &=\ 
\sum_{{n_1\geq0,n_\two>n_1}}
{\frac{{{\left( - \right) }^{\xa(1)}}
     \left( \xa(\two) - \xa(1) -1\right) !}{\left( \xa(\two) -
        3\xa(1) \right) !{{\xa(1)!}^2}\xa(\two)}}\ 
 {z_1}^{n_1}\,{z_2}^{n_{\two}}\cr
&=\ \hskip 12pt
\sum_{n_1,n_2} N^{(2)}_{n_1,n_2}Li_2(e^{2\pi i(n_1t_1+n_2t_2)})\ .
}
}
For explicit values for some of the disk instanton coefficients $N^{(2)}_{n_1,n_2}$,
see e.g.~\doubref\AVii\PM.

In the above basis for the chiral ring, adapted to the special
coordinates $t_A$ on $\MM$, the relative period matrix 
$\Pi^\Sigma_I(t_A)$ takes the form
\eqn\Pexample{
\eqalign{
&\Pi^\Sigma_I(t_A)\ =\ \langle\Gamma^\Sigma,\pi_I(t_A)\rangle\ =\
\pmatrix{ 
1 &\xo & \xs &\del_{\xo}\cF & \cW
\cr
0 &1  & 0 &\del_{\xo}^2\cF & \del_{\xo}\cW
\cr
0 &0 & 1 & 0  & \del_{\xs}\cW
\cr
0 &0  & 0 & 1 & 0
\cr
0 &0  & 0 &0 & 1
\cr
}\, ,
}}
where $\Ga^\Si$ are 3-cycles in the flux sector, 
$\Si\in\{1,2,4\}$, and 3-chains 
in the brane sector, $\Si\in\{3,5\}$. This
matches the form \pmii\ upon setting $\cW_1=\del_{\xo}\cF$,
$\cW_2=\cW$; the diagonal terms can be restored by undoing the
rescalings by $f,g$.

The relative period
matrix \Pexample\ has been previously computed in \PM, as the 
{\it ordinary} period matrix on the absolute cohomology $H^4(M)$ 
of a certain \CY 4-fold $M$ that is related,
via an open-closed string duality, to the D-brane geometry on 
$\cx O(-3)_{\IP^2}$. In particular,
the holomorphic $\cx N=1$ special geometry of the open-closed
string type II string is in this case identical to the special geometry of
the moduli space of the \CY 4-fold $M$. 

To conclude this section, we note that
it should be possible to interpret the open-closed chiral ring
structure constants  \ringele\ in terms of correlation functions
of a boundary TFT. However,
this goal is hampered in the present example for practical 
reasons, namely by the non-compactness
of both the target space and the D-brane world-volumes.
At any rate, the fact that 
$\big(C_i(t)C_j(t)\big)_1^{\ \rho_{op-cl}}=\p_{t_i}\p_{t_j}\cW(t)$
is highly reminiscent of the well-known relation
$\big(C_i(t)C_j(t)C_k(t)\big)_1^{\ \rho_{cl}}=\p_{t_i}\p_{t_j}\p_{t_k}\cF(t)$
in the bulk TFT (here $\rho_{...}$ 
refers to the top elements of the open-closed
and closed chiral rings, respectively).

\subsec{Families of hypersurfaces of degree $>1$}

As discussed in sect.\ 5, the instanton expansion of the 
superpotential can be obtained readily from the 
system of differential equations given in eqs.\egkzf,
for all integers $\nu$ in \moddefii.
Below we sketch briefly the effectiveness of this type of computations
for some hypersurfaces of degree larger then one.

\noi Defining
\eqn\moridd{
\ll1 = (-3,1,1,1,0,0),\qquad \ll2 = (1,-\nu-1,\nu,0,1,-1),
}
the moduli $z_A$ and the GKZ system for the relative cohomology
on the hypersurface $\Z$ follow from the standard
toric definitions \egkz\ and \defza\ for all $\nu$, upon setting
$a_4=a_5=1$.\foot{The vectors \moridd\ can also be
associated to the ordinary Hodge variation on the middle
cohomology $H^4(M)$ on a \CY 4-fold $M$; see \PM\ for details.}
Specifically, the GKZ system \egkzf\ associated with 
\moridd,
\eqn\nuLops{
\eqalign{
{\cal L}_1(\nu)&\ =\ 
{{\xt(1)}}
\big( \xt(1) -(\nu\!+\!1)\,\xt(\two) \big) 
\big( \xt(1)\! +\!\nu\, \xt(\two) \big) +
  \zo\big( 3\xt(1)\! -\! \xt(\two) \big)
   \big( 1 \!+ \!3\xt(1) \!-\! \xt(\two) \big)
   \big( 2 \!+ \!3\xt(1) \!- \!\xt(\two) \big)\ ,
\cr
\widehat{\cal L}_1(\nu)&\ =\ 
\big( 3\xt(1)\! - \!\xt(\two) \big)
\prod_{\ell=0}^{\nu-1}\big( \xt(1) \!+ \!\nu\,\xt(\two)\!-\!\ell \big) -\zt\prod_{\ell=0}^{\nu}\big( \xt(1) \!-\! 
(\nu\!+\!1)\,\xt(\two)\!-\!\ell \big)\ ,
}}
leads to a quick determination of the instanton corrected
superpotential in terms of hypergeometric series  
(for simplicity we consider only the case
$\nu\geq 0$; $\nu<0$ leads to very similar expressions).
The solutions with single logarithmic leading behavior in the $z_A$ are
still given by \mirrormap\ and thus the $\cx N=1$ mirror map does not 
depend on the integer $\nu$. This is a special property of 
the present example which is not true in general.

The instanton expansion of the superpotential is obtained,
as in \doubref\PM\LM, from the solution to \nuLops\ with double logarithmic leading
behavior. The leading logarithmic piece proportional to $(\ln z_1 +3\ln z_2)^2$
is fixed by the solution for $\nu=0$ (c.f., \ThW), 
and its presence reflects the fact that we
considered the cohomology with compact support. Thus our computation
of the superpotential 
differs by boundary terms at infinity from the computation 
for the supersymmetric,
non-compact branes performed in \AVi. The leading logarithms
(as well as more generally terms depending only on the closed string moduli)
can be subtracted by adding boundary terms to obtain the superpotential 
for a supersymmetric brane, which is entirely given by the instanton part.
Alternatively,
one may add inhomogeneous pieces to the 
differential operators \nuLops\ that reflect
the boundary terms at infinity. This is the same 
type of subtraction as that made in \AVi\ and it leads to a solution with the
same instanton expansion but without logarithmic terms. 
At any rate, the instanton contribution to the superpotential
is given for general $\nu$ 
by the following generalized hypergeometric series 
\eqn\niusoopa{
\cx W(\nu) = \sum_{n_1,n_2} {
(-)^{n_1+n_2\nu}\, \Gamma(-n_1+(\nu+1)\, n_2)\Gamma(n_2) \over
\Gamma(-3n_1+n_2+1)\Gamma(n_1+\nu n_2 +1)\Gamma(n_1+1)\Gamma(n_2+1)}
z_1^{n_1}z_2^{n_2}.
}
Inserting the inverted mirror map into $\cx W(\nu)$ then leads to the instanton expansion\Winstexpan\
of the superpotential  in the topological flat coordinates $t_A$. 
We have collected some coefficients $N_{k_1,k_2}$ for
a few choices for $\nu$ in the following tables. The coefficients are 
all integral, as predicted in \OV, 
and moreover agree with the results of \yaip\ obtained by 
a localization computation in the A-model.
\

\vbox{
\def\ss#1{{\scriptstyle #1}}
\def\ps{{\phantom \int}}
$$\hskip -10pt
\def\nual(#1){\noalign{\hrule}&&&&& &&&&&    \nu=#1\cr\noalign{\hrule}}
\vbox{\offinterlineskip\tabskip=-3pt\halign{
%\strut\vrule
\hfil~$\ss{#}\ps$&\hfil~$\ss{#}$&\hfil~$\ss{#}$
&\hfil~$\ss{#}$&\hfil~$\ss{#}$&\hfil~$\ss{#}$
&\hfil~$\ss{#}$&\hfil~$\ss{#}$&\hfil~$\ss{#}$
&\hfil~$\ss{#}$&\hfil~$\ss{#}$&\hfil~$\ss{#}$
%\vrule
\cr
\nual(1)
1& 0& 0& 0& 0& 0& 0& 0& 0& 0& 0\cr 
    -2& 3& -7& 19& -56& 174& -561& 1859& -6292& 21658& -75582\cr 
    5& -10& 30& -100& 350& -1274& 4741& -17844& 67704& -258556& 992256\cr  -32& 63& -182& 620& -2296& 8907& -35486& 143572& -586110& 
    2404978& -9895632\cr 
    286& -545& 1530& -5116& 18816& -73242& 295625& -1222080& 5131594& 
    -21766158& 92908536\cr 
    -3038& 5655& -15600& 51249& -185510& 713610& -2859758& 11793782& 
    -49646532& 212102095& -915808632\cr
\noalign{\hrule}}}
$$}
\vskip-15pt
\vbox{
\def\ss#1{{\scriptstyle #1}}
\def\ps{{\phantom \int}}
$$\hskip 5pt
\def\nual(#1){\noalign{\hrule}&&&&& &&&&   \nu=#1\cr\noalign{\hrule}}
\vbox{\offinterlineskip\tabskip=-3pt\halign{
%\strut\vrule
\hfil~$\ss{#}\ps$&\hfil~$\ss{#}$&\hfil~$\ss{#}$
&\hfil~$\ss{#}$&\hfil~$\ss{#}$&\hfil~$\ss{#}$
&\hfil~$\ss{#}$&\hfil~$\ss{#}$&\hfil~$\ss{#}$
&\hfil~$\ss{#}$&\hfil~$\ss{#}$
%\vrule
\cr
\nual(2)
-1& 0& 0& 0& 0& 0& 0& 0& 0& 0\cr 
    2& 5& 19& 85& 416& 2156& 11628& 64581& 366850& 2121405\cr 
    -5& -18& -84& -460& -2665& -16036& -98583& -615436& -3883803& -24712454
\cr  32& 105& 510& 2900& 17836& 114591& 755310& 5056812& 
    34187916& 232578905\cr 
    -286& -905& -4284& -23980& -147056& -952252& -6384435& -43787296& 
    -304865506& -2143839945\cr 
    3038& 9425& 43680& 240295& 1452830& 9317700& 62170632& 426440826& 
    2983278342& 21168794600\cr
\noalign{\hrule}}}
$$}
\vskip-15pt
\vbox{
\def\ss#1{{\scriptstyle #1}}
\def\ps{{\phantom \int}}
$$\hskip -0pt
\def\nual(#1){\noalign{\hrule}&&&&& &&   \nu=#1\cr\noalign{\hrule}}
\vbox{\offinterlineskip\tabskip=-3pt\halign{
%\strut\vrule
\hfil~$\ss{#}\ps$&\hfil~$\ss{#}$&\hfil~$\ss{#}$
&\hfil~$\ss{#}$&\hfil~$\ss{#}$&\hfil~$\ss{#}$
&\hfil~$\ss{#}$&\hfil~$\ss{#}$&\hfil~$\ss{#}$
&\hfil~$\ss{#}$
%\vrule
\cr
\nual(13)
1& 0& 0& 0& 0& 0& 0& 0\cr 
    -2& 27& -547& 13131& -346256& 9692892& -282809528& 8506210779
\cr  5& -94& 2460& -72180& 2249525& -72674420& 2403644343& 
    -80816694516\cr 
    -32& 567& -14942& 459180& -15221596& 526243689& -18654989678& 
    671776623828\cr 
    286& -4895& 125460& -3801204& 125909616& -4398712716& 158946297875& 
    -5871829419360\cr 
    -3038& 50895& -1279200& 38097081& -1245280910& 43136402580& 
    -1553377686384& 57471237302598\cr
\noalign{\hrule}}}
$$}
\vskip-15pt
\vbox{\vbox{
\def\ss#1{{\scriptstyle #1}}
\def\ps{{\phantom \int}}
$$\hskip 4pt
\def\nual(#1){\noalign{\hrule}&&&&&  \nu=#1\cr\noalign{\hrule}}
\vbox{\offinterlineskip\tabskip=-3pt\halign{
%\strut\vrule
\hfil~$\ss{#}\ps$&\hfil~$\ss{#}$&\hfil~$\ss{#}$
&\hfil~$\ss{#}$&\hfil~$\ss{#}$&\hfil~$\ss{#}$
&\hfil~$\ss{#}$&\hfil~$\ss{#}$&\hfil~$\ss{#}$
&\hfil~$\ss{#}$
%\vrule
\cr
\nual(1712)
-1& 0& 0& 0& 0& 0\cr  2& 3425& 8797969& 26784928225& 89588261367041& 
    318131499524284985\cr  -5& -11988& -39590859& -147317100100& 
    -582323680556665& -2385986201234773711\cr 
    32& 71925& 240477810& 937472436500& 3941883280200586& 
    17285143983849650460\cr  -286& -620780& -2019133809& -7760932508800& 
    -32610125084743556& -144511224912780917884\cr 
    3038& 6456125& 20587246680& 77783427082075& 322535637493687955& 
    1417275737631184228875\cr
\noalign{\hrule}}}
$$}

\vbox{\leftskip 2pc\rightskip 2pc
\noindent{\ninepoint
{\bf Table:} Low degree invariants for various choices of framing
$\nu$. The 
degrees $k_1\geq 0$ in the closed string variable and $k_2\geq 1$ in the
open string variable are listed in the vertical and horizontal directions,
respectively. Sign conventions are as in \yaip.
}}}

\vskip 3mm 
\noi {\bf Acknowledgments:} 
We thank Chris Peters and Christian
R\"omelsberger for comments.
This work was supported in part by the DFG
and by funds provided by the DOE under grant number DE-FG03-84ER-40168. 
WL would like to thank the USC-CIT center in Los Angeles
for hospitality; moreover
WL and NW would like to thank 
the Isaac Newton Institute in Cambridge for hospitality.

\listrefs
\lefin